\documentclass[Chicago]{fsuthesis}
\usepackage{epsfig,contigrefs,axodraw_2}
\def\tp0{^3P_0}
\def\beq{\begin{equation}}
\def\eeq{\end{equation}}
\def\beqa{\begin{eqnarray}}
\def\eeqa{\end{eqnarray}}

\def\lpmb#1{\mbox{\boldmath $#1$}}
\def\half{{\textstyle{1\over2}}}
\def\thalf{{\textstyle{3\over2}}}
\def\fhalf{{\textstyle{5\over2}}}

\def\>{\rangle}
\def\<{\langle}
\def\pr#1#2#3{ {Phys. Rev.\/} {\bf#1}, #2 (#3)}

\def\np#1#2#3{ {Nucl. Phys.\/} {\bf#1}, #2 (#3)}

\def\pl#1#2#3{ {Phys. Lett.\/} {\bf#1}, #2 (#3)}

\def\zp#1#2#3{ {Z. Phys. }{\bf#1}, #2 (#3)}
\def\ptp#1#2#3{ {Prog. Th. Phys. }{\bf#1}, #2 (#3)}
\def\fbs#1#2#3{ {Few Body Syst. }{\bf#1}, #2 (#3)}

\fourlevels
\title{EFFECTS OF BARYON-MESON INTERMEDIATE STATES ON BARYON MASSES}
\author{Danielle Morel}
\uauthor{DANIELLE MOREL}
\thesistype{dissertation}
\department{Department of Physics}

\headofdept{Kirby W.\ Kemper}
\deptheadtitle{Chairman}
\deanofschool{Donald J. Foss}
\degree{Doctor of Philosophy}
\gradyear{2002}
\dept{Physics}
\majorprof{Simon Capstick}
\majorprofdegree{Ph.D.}
\outcommmember{Gregory Riccardi}
\commmemberc{Jorge Piekarewicz}
\commmemberb{Paul Eugenio}
\commmembera{Howie Baer}
\semester{Spring}
\college{Arts and Sciences}
\defensedate{January 23, 2002}
\begin{document}
\dedication{To my family}
\def\acknowledgementtext{
\hskip\parindent
This dissertation could not have been achieved without the support,
guidance, and infinite patience of Professor Simon Capstick. I will
always be grateful to him for seeing in me the potential to undertake
and, despite some rough times, complete this project. I would also
like to extend my heartfelt appreciation to Professor Winston Roberts
of Old Dominion University, for his help and support through the last
year of this project, and to Professor Jorge Piekarewicz of Florida
State University, for his keen insights, and unwavering good
spirits. Many thanks to Professors Baer, Eugenio, and Riccardi for
reviewing this manuscript and offering precious advice. Finally I
would like to thank Professors Dennis and Riccardi for giving me
access to the computing power of the FSU Department of Physics
Computer Cluster without which this project would not have been
feasible, and the FSU College of Arts and Sciences for providing the
funding necessary to make the cluster a reality.

This work was supported in part by the U.S.~Department of Energy under
Contract DE-FG02-86ER40273.
}
\def\abstracttext{%
The mass shifts of experimentally well-known baryons due to
meson-baryon self-energy loops are calculated, and their impact on the
observed splitting of the baryon spectrum is
studied. Configuration-mixed wave functions adapted from
a `relativized' model are used with the $\tp0$ model to provide
predictions for the strength and analytical momentum dependence of the
strong vertices. Intermediate states include all the lightest
pseudoscalar and vector mesons and corresponding baryons required to
provide a complete set of spin-flavor symmetry related baryon-meson
states. The sum over intermediate-state baryons is extended to include
the second ($N=3$) band of negative-parity excited states, to provide
the most complete calculation of its kind to date.

It is found that with reduced-strength one-gluon-exchange interactions
between the quarks, roughly half of the splitting between the nucleon
and Delta ground states arises from loop effects. The effects of such
loops on the spectrum of negative-parity excited states are also
studied, and it is found that the resulting splittings are sensitive to
configuration mixing caused by the residual interactions. With the
extensive set of intermediate baryon-meson states used, a reasonable
correspondence is found between model masses and the bare masses
required to fit the masses of the states extracted from data analyses.
}
\frontmatterformat
\titlepage
\maketitle
\dedicationpage
\prefacesection{Acknowledgements}\acknowledgementtext
\tableofcontents
\listoftables
\listoffigures
\newpage
\abstractsection{ABSTRACT}
\maintext
\chapter{Introduction}
\label{intro}
\pagenumbering{arabic}

The field of nuclear physics spans a very wide range of phenomena and,
despite a long history, is still riddled with a large number of
unanswered questions. Within it, the field of hadron physics finds
itself in a `bridge' position between the visible world of atoms where
electrons and nuclei rule, and the realm of high energy physics where
quarks, gluons, and other such `invisible' elementary forms of matter
are the focus of attention. Hadrons are ultimately what our world is
made of, {\it i.e.}~the fundamental pieces of matter bound
together to become the neutrons and protons (two examples of hadrons),
then further combined to form the nucleus of each atom charted on the
periodic table of elements.

Studying hadrons therefore means trying to understand how and why
particles like quarks and gluons are combined to produce the
experimentally observed spectra of objects labeled `mesons'
(predominantly quark + anti-quark + gluons states, such as pions) and
`baryons' (predominantly three-quark + gluons states such, as protons
and neutrons) as opposed to any other possible
combination. Additionally, while an electron can be removed from an
atom and isolated to study how forces act upon it, the `strong force'
that holds quarks together is so strong that we have not, and
according to our current understanding will not, see a single quark in
isolation. Matter can be disintegrated down to its quark constituents
in highly energetic accelerators but the length of time quarks remain
individual particles (i.e. not bound to others) is incredibly short,
and ultimately they end up bound to each other in hadrons. Therefore
much about the strong force and how it binds quarks into hadrons
remains poorly understood.

Not only is the study of the strong interactions pushing the limit of
current experimental facilities, but theoretical tools that are known
to work well with electromagnetic or high energy phenomena have very
limited use in the study of quarks and gluons, at least at the energy
scale where they form bound states such as baryons and mesons. In
addition, the theory which is now accepted to be correct for strong
interaction physics is so complex that the equations governing it can
only be solved exactly for a very limited number of cases, leaving
sizeable gaps in our understanding.

Theoretical work is therefore ongoing in two main areas. One is to
broaden the range of applicability of exact solutions of the theory of
strong interactions via novel computing techniques and algorithms. The
other is to develop and/or improve models which, although not exact,
give a good overall picture of many manifestations of the strong force
and open windows to the physics behind many experimentally observed
phenomena. Within this context, the broad goal of the work presented
in this thesis is to remove a level of approximation in one already
successful baryon spectroscopy model, {\it i.e.}~a model which
describes and predicts the number and properties of baryon states that
should be `seen' experimentally. Doing so should improve the ability
of this model to explain some of the intricacies of the baryon
spectrum, while increasing its predictive power and helping resolve
some of the discrepancies between predictions and observations.
\section{Quantum Chromodynamics}
It is now accepted that the building blocks of matter are the quarks,
leptons, and gauge bosons. Quarks come in six different species or
`flavors', and are known as the up `$u$', down `$d$', strange `$s$',
charm `$c$', top `$t$', and bottom `$b$' quarks. The six flavors of
leptons are the electron `$e^-$', muon `$\mu^-$', and tau `$\tau^-$'
and their associated neutrinos $\nu_{e}$, $\nu_{\mu}$, and
$\nu_{\tau}$. Quarks and leptons form a group of particles called
fermions. Each has half-integer spin and an associated anti-particle,
for example the positron `$e^+$' or the anti-top quark `$\bar{t}$',
which are identical in mass but different in charge and color (another
quark quantum number discussed more below) to their particles. The
gauge bosons are integer-spin particles and are the photon `$\gamma$',
$Z, W^{\pm}$, gluon `$g$', and the proposed graviton.

There are four fundamental interactions, each
with one or more particles associated with it that carry the
force. {\it Gravity}, the weakest of these interactions, is thought to
be mediated by massless, spin-two gravitons. The {\it weak
interaction}, whose force is carried by spin-one, self-interacting $Z$
and $W^{\pm}$ bosons, is really a component of the electro-weak force and is
responsible for radioactive beta decay processes. The other part of
this force manifests itself via {\it electromagnetic interactions},
which act between electrically charged objects and are mediated by
spin-one, electrically neutral photons that do not interact among
themselves. Finally, the {\it strong interaction}, which acts between
color-charged objects, is mediated by self-interacting vector bosons
called gluons and is responsible for nuclear binding and the
interactions of the constituents of the nuclei. The quarks, which come
in three colors and have fractional electric charge, are known to have
strong, electromagnetic, weak, and gravitational interactions. The
leptons, such as the electron, are subject to all forces except the
strong one, as they do not carry a color charge, while the neutrinos
have neither strong nor electromagnetic interactions.

The strong force is `weak' at very short distances, creating what is
called asymptotic freedom (quarks appear to behave like free
particles), but grows infinitely strong at `large' distances,
resulting in the phenomenon known as confinement (a quark cannot be
isolated like an electron or a proton).  The strong interactions
provide the `glue' to hold the quarks together to make hadrons, the
strongly interacting particles, the mesons and baryons. Although the
number of quarks within hadrons is not defined by quantum field
theories, several models have baryons composed of three `constituent'
quarks, which have half-integer spin and obey Fermi-Dirac
statistics. Mesons on the other hand are described as integer spin
quark-antiquark pairs that obey Bose-Einstein statistics. To be
observable, particles such as hadrons are required by confinement to
be colorless objects (color singlets). Baryons must therefore have one
red, one blue, and one yellow quark (as $RBY$ = white) and mesons are
allowed to be a colorless combination of $R\bar{R}$, $B\bar{B}$, and
$Y\bar{Y}$ quark-antiquark pairs. There is growing evidence for the
existence of other quark and gluon states such as glueballs (pure
gluon states) and hybrids ({\it eg.} $qqqg$), as well as hypothesized states
such as diquonia ($q{\bar{q}}q{\bar{q}}$), dibaryons ($qqqqqq$), and
others.

The equations describing the electromagnetic interactions were
formulated by Maxwell and form the basis of Quantum Electrodynamics
(QED). Correct theories for weak and strong interactions came later.
It is now widely accepted that a field theory equivalent to QED exists
for strongly interacting particles and is known as Quantum
Chromodynamics (QCD), with gluons mediating the forces between colored
quarks, which are analogous to photons. The QCD Lagrangian has the
form
\begin{equation}
{\cal L} = -\frac{1}{4}F_{\mu\nu a} F_a^{\mu\nu} +
{\bar{\Psi}}(i{\rlap D / }-m)\Psi,
\end{equation}
where $D_{\mu}=\partial_{\mu}-igA_{\mu}^at^a$ is the covariant
derivative, $F_a^{\mu\nu}= \partial^{\mu} A_a^{\nu} - \partial^{\nu}
A_a^{\mu} + g\sum f_{abc} A_b^{\mu} A_c^{\nu}$ is the field tensor,
and $A_a^{\mu}$ are the gluon fields ($a = 1~{\rm to}~8$). The last
term in the field tensor definition indicates that, unlike the photons
of QED, gluons interact with each other, giving QCD its non-abelian
behavior (for more information on QCD see for example
Refs.~\cite{Yn:1999,GSS:2000,Gr:1999}). Unfortunately, unlike QED,
there is as yet no obviously successful way to go from the QCD
Lagrangian to a complete understanding of the large number of observed
hadrons and their properties. Lattice QCD, a field theory that
replaces space-time with a lattice of discretely spaced points (colored
sources -quarks- at the junctions and color electric flux lines
-mediated by gluons- as links between them), is making visible
progress towards that goal by using numerical techniques to solve
otherwise unreachable problems, but calculations beyond masses and
static properties of the ground states and the lightest negative
parity excited baryon states are still in the future. Other methods,
such as expansions based on the large $N_c$ (number of colors) limit
of QCD, and effective field theories (theories that replace part of
unknown interactions by physically sound approximations), are limited
in their scope and are currently unable to provide the global yet
detailed understanding needed for a description of all aspects of
strong interaction physics.
\section{Non-perturbative QCD and QCD-based Models}
It is relatively easy to treat electroweak interactions via
perturbation theory (PT), but strong interactions of hadrons involve
dealing with QCD, where one cannot expect much of PT in a situation
that is fundamentally not one of weak coupling, as is explained below.

\begin{figure}
\begin{center}
\epsfig{file=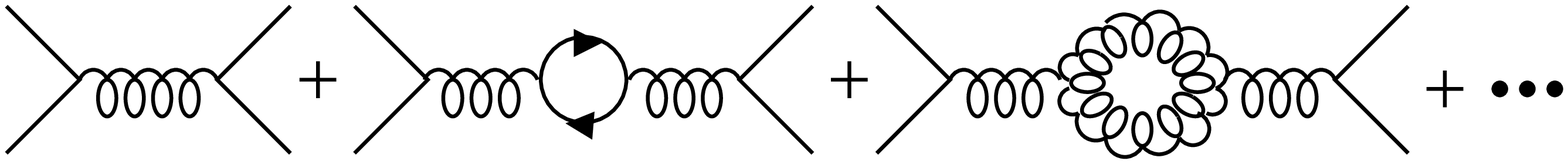,width=5.75in,height=0.75in,angle=0}
\caption[Strong coupling constant]{\label{Strong_a}Perturbative
expansion of the gluon-gluon interactions defining the strong coupling
constant $\alpha_s(Q^2)$.}
\end{center}
\end{figure}

The usual way to treat local interactions is through PT, {\it i.e.}~by
expanding various quantities in powers of the coupling constant. In
QED, it is useful to define an effective coupling constant
$\alpha(Q^2)$, which gives the momentum transfer ($Q^2$) dependence of
the renormalized vertex function. This function receives contributions
from vacuum polarization graphs which describe the electron loop
corrections to the photon propagator. The result is an effective coupling
$\alpha(Q^2)$ which increases with $Q^2$. In QCD however, this situation is
complicated by the gluon self-interactions (gluons carry color
charge). The effective quark-gluon vertex can be summed over all
orders of the renormalized coupling (see figure~\ref{Strong_a})
and has the form
\begin{equation}
\label{a_s}
\alpha_s(Q^2) = \frac{12\pi}{(33 - 2 N_f)ln(Q^2/\Lambda_{QCD}^2)}.
\end{equation}
where $N_f$ is the number of flavors, and $\Lambda_{\rm QCD}$ is the
momentum scale at which $\alpha_s$ becomes strong as $Q^2$ is
decreased. 

From Eq.~\ref{a_s} it can be seen that at large $Q^2$, or short
distances (relative to the $\Lambda_{\rm QCD}$ scale and for $N_f\leq
16$), $\alpha_s(Q^2)\rightarrow 0$ (a property known as asymptotic
freedom) so that hard processes (processes calculable using
perturbative QCD as the momentum transfer is large enough to produce a
small value of $\alpha_s$) such as deep inelastic scattering, some
weak decays of heavy flavors, and some experimentally seen but
theoretically highly suppressed strong decays of heavy quarkonia, can
successfully be treated perturbatively. On the other hand, at small
momentum transfers, or large distances, $\alpha_s(Q^2)$ grows quite
large and PT becomes invalid. Unfortunately, this is the region
relevant to strong decays of hadrons composed of light quarks,
electromagnetic transitions, and the weak decays of hadrons containing
light flavors, like hyperons (baryons containing one or more strange
quarks) or kaons (mesons containing one strange quark or
anti-quark). Experimental measurements yield a value of $\Lambda_{\rm
QCD} \approx 200$ MeV. The strong interactions become strong at
distances larger than $\sim 1/\Lambda \approx 1$~fm, which is roughly
the size of the light hadrons.

Within QCD, there are two main phenomena that are essentially
non-perturbative and therefore cannot be obtained even by summing
entire perturbative series. The first is confinement, and the second
is the dynamical breaking of chiral symmetry. Confinement is related
to the interaction energy, which increases with distance in contrast
to the Coulomb energy. The breaking of chiral symmetry gives a
dynamical mass of several hundred MeV to light quarks, which are known
to have current-quark masses of only a few MeV at large momentum
scales. Both phenomena are connected to non-zero vacuum expectation
values, which would vanish at any order of PT: the so-called quark
($<q\bar{q}>$) and gluon ($<gg>$) condensates.

There are two main alternatives to PT based on field theory. One is
the QCD duality sum rules approach, which uses a short distance
expansion of products of operators. This method allows certain
quantities to be calculated in terms of the condensates $<q\bar{q}>$,
$<gg>$, and perturbatively calculable coefficients. These quantities
can also be calculated in terms of hadron properties such as masses,
widths, and branching ratios, hence the duality. The other is lattice
QCD, which attempts to completely calculate hadronic properties from
first principles (the QCD Lagrangian on a lattice of space-time points
based on the work of Wilson~\cite{Wi:1974}, see Ref.~\cite{Cr:1983}
for an introduction to lattice QCD) based on the
physical idea that the long-distance properties of QCD are the most
important (confinement and dynamical breaking of chiral symmetry), and
that short distance properties can be reached by extrapolation. These
techniques permit strong coupling calculations based on the assumption
that the unrenormalized coupling constant is very large, which use
Monte Carlo simulations based on Feynman path-integrals.

Recently a covariant approach to the description of hadron structure
has been developed, based on the Bethe-Salpeter equation and the
Schwinger-Dyson method of solving field theory. This approach has been
widely developed for the description of meson, but so far restricted
to the descriptions of ground-state baryons.

Although these methods are increasingly useful, they remain
technically very complex and somewhat limited in their
applicability. Additionally, using these techniques, it is often
difficult to obtain physical insight (especially on the lattice) about
phenomena such as decay mechanisms or confinement (although
confinement is a natural consequence of the lattice, {\it i.e.}~the
area of the Wilson loop gives an energy $\propto r$ so produces a
linearly rising potential, it is not a proof of its existence). This
highlights the continued need for other methods which are inspired by
QCD but not necessarily derived from it. The inability to calculate
with QCD in the low $Q^2$ regime has made it necessary to use
phenomenological models of hadron structure based on expectations of
the low energy behavior of QCD. The quark potential model and other
dynamical models were created to fulfill this purpose.

Within the framework of the quark potential model, the spectra of
mesons and baryons, as well as their strong, weak, and electromagnetic
decays have been successfully calculated. This model allows for direct
calculation of relevant matrix elements for each definite
decay and provides transparent direct links to experimental
data. Its simplicity implies a lack of theoretical foundation in
QCD, but despite that fact its past and present empirical successes
are impressive.
\section{Corrections to the Quark Model}
In QCD there are $qqq(q\bar{q})$ configurations possible in baryons,
and these must have an effect on the constituent quark model, similar
to the effect of unquenching lattice QCD calculations. These effects
can be modeled by allowing baryons to include baryon-meson (BM)
intermediate states, which lead to baryon self energies and mixings of
baryons of the same quantum numbers. A calculation of these effects
requires a model of baryon-baryon-meson (BB$^\prime$M) vertices and
their momentum dependence. It is also necessary to have a model of the
spectrum and structure of baryon states, including states not seen in
analyses of experimental data, in order to provide wave functions for
calculating the vertices, and to know the thresholds associated with
intermediate states containing missing baryons.

The goal of the present work is to self-consistently calculate such
effects for a set of experimentally well known baryon states. The
method and results are presented as follows. In Chapter~\ref{loops}
the work of several authors who have made contribution to this field
is reviewed, and important elements are extracted and related to the
present work. The nature and extent of the present research is then
discussed in more detail, highlighting the improvements needed to be
made to this type of calculation. Chapters~\ref{QuarkModel}
and~\ref{StrongDecays} present an overview of the different methods
adopted for use in this research. Results are then presented in detail
in Chapter~\ref{ResI}, and finally Chapter~\ref{Concl} offers
conclusions and outlook for future extensions of this work.
\chapter{Baryon Self Energies}
\label{loops}
\begin{figure}
\begin{center}
\epsfig{file=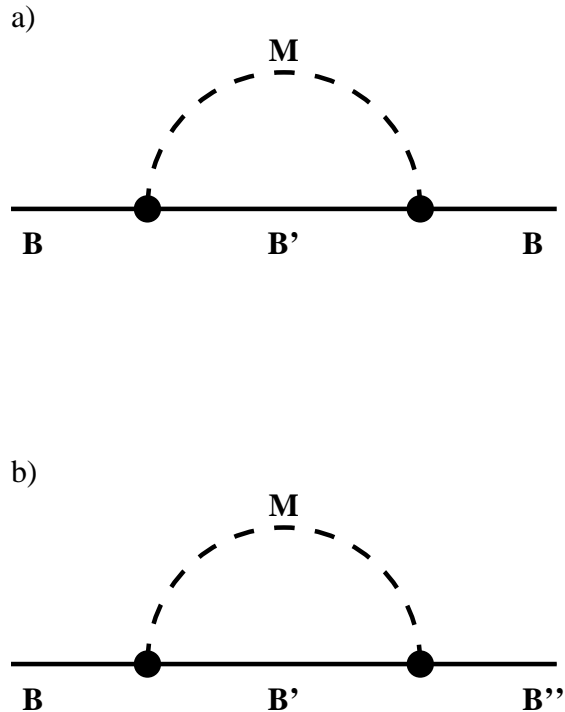,width=3.2in,height=4in,angle=0}
\caption[Meson loop contribution to the self energy and mixing of a
baryon]{\label{mixing}Contributions from a baryon-meson loop to a) the self
energy of baryon $B$ and b) the mixing of baryons with same quantum
numbers. The dashed arc represents a virtual meson.}
\end{center}
\end{figure}

Baryon self energies due to BM intermediate states and BM decay widths
can be found from the real and imaginary parts of loop diagrams (see
Figure~\ref{mixing}). The size of such self energies can be expected to
be comparable to baryon widths. For this reason, they cannot be
ignored when comparing the predictions of any quark model with the
results of analyses of experiments. Since the splittings between
states which result from differences in self energies can be expected
to be comparable to those that arise from the residual interactions
between the quarks, a self-consistent calculation of the spectrum
needs to adjust the residual interactions, and with them the wave
functions of the states used to calculate the BB$^\prime$M vertices,
to account for these additional splittings.

Earlier studies have brought these facts to the attention of the
nuclear physics community, each highlighting different aspects of the
problem. What seems to be missing is a consistent and complete
calculation of the effects of the self energies on the baryon
spectrum. By looking back at the existing literature, lessons can be
learned on how to accomplish such a task as thoroughly as possible.
%
\section{Existing Work}
Previous calculations of the self energies of ground state and
negative-parity excited baryons use baryon-meson intermediate states
consisting of ground states. The work of Zenczykowksi~\cite{Ze:1986}
takes the point of view that the `residual' interquark interactions
are unimportant, and that hadronic loop effects dominate the observed
splitting and mixing pattern of the ground and first (negative-parity)
excited states of baryon states in the octet and decuplet SU(3)$_f$
representations.

This work claims that at least two thirds of the observed splittings
in these states can be attributed to such effects, as can the mixing
angles between states due to these effects. In a simplifying limit, a
formula relating the $\Sigma-\Lambda$ and $\Delta-N$ mass differences,
derived using one-gluon exchange by de Rujula, Georgi, and
Glashow~\cite{DeRGG:1975}, can be attributed to the effects of
hadronic loops.

This calculation uses only spatial ground state intermediate baryons,
but unlike those of some other authors, considers a complete set of
accessible SU(3)$_f$ intermediate states. This means that for $N$ and
$\Delta$ baryons the intermediate states with pseudoscalar mesons were
$N\pi$, $\Delta\pi$, $N\eta$, $\Delta\eta$, $N\eta^\prime$,
$\Delta\eta^\prime$, $\Sigma K$, $\Lambda K$, $\Sigma^* K$, and those
with vector mesons were $N\rho$, $\Delta\rho$, $N\omega$,
$\Delta\omega$, $\Sigma K^*$, $\Lambda K^*$, and $\Sigma^* K^*$. This
calculation also considered the self energies of strange baryons,
where a different (and larger) set of intermediate states is
possible. A dispersion integral relates the shift in the squared mass
of a baryon state to properties of the intermediate states and a
spectral function $\rho(s,m_B^\prime,m_M)$, which depends on the
nature of the baryons $B$ (initial) and $B^\prime$ (intermediate)
involved in the loop diagram,
\begin{equation}
m_B^2-(m_B^0)^2=\sum_i w_i^B \int_{s^{(i)}_{\rm thr}}
{\rho(s,m_B^\prime,m_M)\over m_B^2 - s} ds.
\end{equation}
Here the sum runs over all (open and closed) decay channels
$i={B^\prime M}$ with $\sqrt{s^{(i)}_{\rm thr}}=m_{B^\prime}+m_M$, and the
weights $w_i^B$ give the spin and flavor [SU(6)] overlaps between the
$B$ and $B^\prime M$ states. For $B$ and $B^\prime$ baryons restricted
to ground states, and without SU(3)$_f$ breaking in these ground-state
baryon wave functions, these weights have the important property that
\begin{equation}
\sum_i w_i^B = 48,
\end{equation}
for all baryons $B$, as long as the sum runs over all intermediate
states $i$ allowed by the quantum numbers. This means that in this
symmetry limit the mass shift due to these effects is the same for all
of the ground state baryons, and so no mass splittings are generated
by loop effects, as might be expected. This observation is critical,
as it makes clear that without the inclusion of at least this set of
intermediate states, calculations of these effects do not start from
the symmetry limit and so cannot be expected to give physically
meaningful results.

Away from this limit the nature of the spectral functions $\rho$
becomes important. Their values were calculated using the $^3P_0$
model with universal radii for the mesons and baryons, taking into
account the spin, flavor, and spatial structure of the baryons and
mesons involved.

The author concludes that, with reasonable hadron radii, about two
thirds of the splittings and mixings in ground and negative-parity
excited state baryons must arise from hadronic loop effects, and that
this allows the use of a significantly smaller coupling constant
$\alpha_s$ in the quark residual interactions which may explain the
rest of the splittings. It is crucial that a `complete' set of
SU(6)-related $B^\prime M$ intermediate states is employed.

The work of Blask, Huber, and Metsch~\cite{BHM:1987} is similar to
that of Zenczkowski, except that vector mesons are not taken into
account in the intermediate states, and the point-like mesons are
coupled to baryon states using an elementary-meson emission (EME)
model with the meson-quark coupling fixed from the $\pi NN$, $K\Lambda
N$ and $\eta NN$ coupling constants derived from analyses of
experimental data. The recoil energy of the intermediate baryon is
neglected, but intermediate state thresholds are given by using the
physical masses of the intermediate hadrons. Baryon masses including
these loop effects are found by diagonalising an effective Hamiltonian
which includes a term for the internal dynamics of the baryon and
meson, and an energy-dependent term
\begin{equation}
\delta H_b(E) = \hat{H}_c {1\over E +i\epsilon - H} \hat{H}_c,
\end{equation}
where $\hat{H}_c$ describes the coupling between baryons and mesons
and $H$ is the full Hamiltonian. 

Similar conclusions about the importance of such effects are made,
although this calculation suffers from a point-like treatment of the
mesons (leading to overestimated widths for decaying baryon states)
and some problems in the resulting spectrum which likely arise from
the restricted set of intermediate states. This calculation also
hints at a possible cancellation between the spin-orbit effects due to
the one-gluon-exchange residual and splittings which arise from the
inclusion of $B^\prime M$ intermediate states.

Brack and Bhaduri~\cite{BrBh:1987} calculate self energies of the
nucleon and $\Delta$ ground states only, using only pions as
intermediate mesons, but do not restrict the intermediate $N$ and
$\Delta$ baryon states to spatial ground states. They find that the
difference in the self energies of the nucleon and $\Delta$ ground
states converges to within 5 MeV of the large $N$ result when a set of
intermediate baryons up to and including the $N=3$ band (second band
of negative-parity orbital excitations) states is employed. They find
that, in their model, the difference in the pionic self energies of
the odd-parity excited states and the ground state converges too
slowly to make definite statements.

Part of this trouble with convergence may be due to their model of the
$BB^{\prime}M$ amplitudes, which simply attaches a pion to the quarks
with a (nonrelativistic) pseudoscalar coupling, with an additional
axial form factor
\begin{equation}
F_\pi({\bf k}^2)=1/(1+{\bf k}^2/\Lambda_{\pi}^2),
\end{equation}
with $\Lambda_{\pi}=1275$ MeV, corresponding to the mass of the $a_1$
meson. Since their loop amplitudes involve elementary intermediate
pions, they include a factor of $1/\omega_k$, where
$\omega_k=\sqrt{{\bf k}^2+m_\pi^2}$ is the pion energy, from the
normalization of the wave function of the intermediate pion. This
factor is not present in the pion center of mass wave function in
nonrelativistic models which treat it as a composite
particle. Although the presence of this factor has the effect of
further suppression of high-momentum contributions to the integral
over the loop momentum, the net result is that it is still likely that
the effective pion-nucleon vertex in this model is too soft. In
subsequent models and the present work a more rapid decrease of the
vertex amplitudes with $k^2$ is shown to produce better results for
the mass shifts, and can be attributed to an effective size for the
operator which creates a constituent quark-antiquark pair (see Geiger
and Isgur~\cite{GeIs:1991-2}).

The intermediate states are described by simple unmixed harmonic
oscillator wave functions. The excitation spectrum of the intermediate
states is taken to be either harmonic oscillator plus zero-range
(contact) spin-spin potential, or the Isgur-Karl potential which is
modified by anharmonicities in the spin-independent potential, which
gives a more realistic spectrum for the energy of the intermediate
states. They show that, at least for the nucleon-$\Delta$ splitting,
the details of this spectrum are unimportant. It can be expected that
they will become very important, however, if this calculation was
extended to the self energies of excited states, as these depend
crucially on the positions of the thresholds due to the opening of
various channels for excited states to decay to excited states.

The important conclusions of this work are: convergence of the
nucleon-$\Delta$ mass difference in the sum over excited intermediate
states can be demonstrated; it is misleading to include only baryon
ground states as intermediate states, as inclusion of excited states
reduces the difference in the nucleon and $\Delta$ self energies
substantially; their final results depend sensitively on the chosen
(axial) radius of the nucleon, as expected, and changing the gluonic
hyperfine splitting changes the difference in the self energies of the
nucleon and $\Delta$; and if the gluonic hyperfine splittings are too large 
($>250$ MeV) it is impossible to fit the observed $\Delta-N$
splitting. It is also noted that one-pion exchange can, with some
adjusted parameters (a reduced strength coupling to the quarks), be
made to simulate these effects. Poor convergence was found in the
calculation of the self energies of the negative-parity excited
states, an issue which will be resolved in the present work.

The work of Horacsek, Iwamura and Nogami~\cite{HoIwNo:1985} would
appear to partly contradict that of Brack and Bhaduri, with
$\Delta-N=20$ MeV from the inclusion of baryon-pion intermediate
states. However, the approximation of each intermediate state quark
moving in a single central potential (shell model) is used, so that
intermediate excited baryon states are described as individual excited
quark substates. This means that the intermediate states are far from
a basis of hadrons, and so this calculation ignores what we know about
the spectrum of confined hadrons in the intermediate state, and the
resulting thresholds. Both of these calculations are incomplete
because they do not include contributions from mesons other than
pions.

Silvestre-Brac and Gignoux~\cite{SBGi:1991} examine the self energies
of only the lowest lying negative-parity excited states, and focus on
total spin 1/2 and 3/2 spin-orbit doublets in the $N$, $\Delta$,
$\Lambda$, and $\Sigma$ flavor sectors. They correctly use a complete
set of SU(6)-related intermediate states, but as in Zenczykowksi's
work, these are restricted to spatial ground states. No configuration
mixing is allowed between states due to interquark Hamiltonian
($H_0$), so spin-orbit partners in these negative-parity excited
states are degenerate. The calculation also uses only one radius for
all baryons and all mesons, {\it i.e.}~uses the simplifying assumption
of SU(3)$_f$ symmetry in the wave functions. Bare masses are solved for
self-consistently, {\it i.e.}~are free parameters. The decay
thresholds are found using physical masses for the intermediate
virtual hadrons, which is equivalent to adopting dressed states in the
propagators of the intermediate hadrons, and this is of course crucial
to obtaining a correct description of widths. The calculation also
uses a cut-off factor in the integrals over the loop momentum. The
authors justify this by comparison to calculations which use
elementary pions in the intermediate state and so have a further
suppression factor of the inverse of the pion energy, $1/\omega_k$,
and from a lack of information about strong vertices at large relative
momenta of the final-state hadrons.

Their conclusions are that hadronic loops are important ingredients in
the understanding of spin-orbit splittings, with a satisfactory
description of the order and magnitude of the spin-orbit splittings of
negative-parity excited baryons resulting from their
calculation. However, this latter conclusion seems premature given
that it has been shown by Brack and Bhaduri~\cite{BrBh:1987} and
Geiger and Isgur~\cite{GeIs:1991-2} that the restriction of the
intermediate state baryons to ground states results in self energies
which have not converged.

The calculation of Fujiwara~\cite{Fu:1993} uses antisymmetrized
$(3q)(q\bar{q})$ cluster-model wave functions composed of simple
harmonic oscillator wave functions and the plane-wave relative motion
to describe the baryon-meson intermediate states. The decay operator
employed is unlike those in other calculations, as pairs are created
by an interaction between a quark and a quark-antiquark pair creation
vertex which is consistent with the residual interactions between the
quarks in the hadrons (see also Ackleh, Barnes and
Swanson~\cite{AcBaSw:1996}). In particular, it contains the contact,
tensor, and spin-orbit interactions arising from one-gluon-exchange
between the quarks. The self energies of ground states and lowest
lying negative-parity excited states of $N$, $\Delta$, $\Lambda$ and
$\Sigma$ baryons are calculated using intermediate states restricted
to ground state pseudoscalar and vector mesons, and ground state octet
and decuplet baryons. The nonrelativistic approximation is made in the
energy denominators in order to allow analytic treatment of the loop
integrals involved in the evaluation of the self energies.

Rather than adopt the momentum dependence of the vertex amplitudes
which arises from the structure of the hadron states, the simplifying
assumption of a universal dependence on the relative momentum $k$ of
the intermediate hadron pair is adopted. As in other calculations,
this dependence is modified from the exp$(-k^2/6\alpha^2)$ dependence
given by a nonrelativistic evaluation of the decay amplitude in the
presence of recoil, where $\alpha$ is the harmonic oscillator size
parameter, in order to further suppress high-$k$ contributions to the
loop integrals. In this case it is simply given the value
exp$(-k^2/3\alpha^2)$. Flavor-symmetry breaking is ignored in the
pair-creation interaction for simplicity. 

The results show that it may be possible to arrange a cancellation
between spin-orbit splittings arising from the interactions between
the quarks and from loop effects, and to describe the mixings and
decay widths of these states in the same model. Notable exceptions are
the flavor singlet (lowest lying) negative-parity $\Lambda$ states
$\Lambda(1405)$ and $\Lambda(1520)$ which are about 100 MeV too heavy,
as in simple three-quark models.

As mentioned previously, conclusions made in the models described
above about spin-orbit forces in negative parity excited baryons are
likely to be premature, given the information provided about
convergence by Brack and Bhaduri~\cite{BrBh:1987}. It is shown in the
present work that the inclusion of negative-parity excited baryons in
the intermediate states and configuration mixing in their wave
functions are crucial to the accurate calculation of mass shifts of
these states.

From the above work it is clear that a self-consistent and successful
model of baryon self energies must employ a complete set of
spin-flavor symmetry related $B^\prime M$ intermediate states, and at
the same time must include excited baryon states up to at least the
$N=3$ band in order for the sum over intermediate states to have
converged. This requires a detailed and universal model which is
capable of relating the baryon spectrum and the decay amplitudes of a
wide variety of baryon states to a wide variety of baryon-meson final
states in an efficient way. It is also clear that it will be necessary
to modify the usual momentum dependence of the decay amplitudes
calculated in this model to take into account the size of the
constituent quark-pair creation vertex.

In addition, the size of these loop effects requires that the
interactions between the quarks required to fit the observed spectrum
be changed by the presence of these loop effects. It is inconsistent
to not then also change the wave functions used to calculate the vertex
amplitudes and to examine the effect of these changes on the
self-energies. Brack and Bhaduri~\cite{BrBh:1987} have shown that the
$\Delta$-nucleon splitting may not be sensitive to such details, but
from the sensitivity to the structure of the interquark Hamiltonian
used to describe the hadron states observed in many of these
calculations, it can be expected that this will be an important effect
in the calculation of the self energies of the negative-parity excited
baryons.
%
\section{Current Work}
Based on all the lessons learned above, our goal is then to calculate
the energy-dependent self energy of baryon $B$ given by
\begin{equation}
\label{loop_eq}
Re[\Sigma_B(E)] = \sum_{B'M} {\cal P}\int_0^\infty \frac{k^2dk
{\cal M}_{BB'M}^{\dagger}(k){\cal M}_{BB'M}(k)}{E-\sqrt{M_{B'}^2 + k^2} -
\sqrt{m_M^2 + k^2}}.
\end{equation}
where ${\cal M}_{BB'M}$ is the analytical strong decay matrix element
of initial baryon $B$ decaying into two hadrons, baryon $B'$ and meson
$M$, as calculated with the $\tp0$ pair creation model. The integral
is taken over the relative momentum $k$ between baryon $B'$ and meson
$M$.

Eq.~\ref{loop_eq} comes from first-order, time-ordered perturbation
theory, and is not a fully relativistic, frame-independent
equation. Here it is evaluated in the rest frame of the decaying
baryon $B$. It has both real and imaginary parts but the real part
only is extracted by taking the principal part of Eq.~\ref{loop_eq},
in effect performing the integration everywhere along the real line
except over a small symmetric interval centered over the pole
location.

We calculate the self energies of Eq.~\ref{loop_eq} for the ground
states Nucleon and $\Delta$, and lightest negative-parity excited
states. As detailed further in appendix~\ref{ap:WF}, baryons have
color, spin, flavor, and spatial wave functions obtained from the
different ways three colored, flavored, spin-1/2 quarks can be
combined with the relative angular momenta of the three-body system's
two relative coordinates (see figure~\ref{figB1}) $\lpmb{\rho}$ and
$\lpmb{\lambda}$. With no orbital angular momentum, {\it i.e.}~${\bf L}={\bf
l}_{\rho} + {\bf l}_{\lambda}=0$ and
$n_{\rho}=n_{\lambda}=0$, only two states are possible:
$L^P = 0^+ \otimes \{ S=\half$ or $S=\thalf \}
\rightarrow J^P = \{ \half^+,\thalf^+ \}$ where the total angular
momentum ${\bf J}={\bf L}+{\bf S}$ and parity $P=(-1)^{{\bf l}_{\rho}
+ {\bf l}_{\lambda}}$. The lightest of these states is the Nucleon,
with $J^P=\half^+$ and flavor wave function either $uud$ (proton) or
$udd$ (neutron). The other combination gives us the $\Delta$ with
$J^P=\thalf^+$ and flavor wave function $uuu$ ($\Delta^{++}$), $uud$
($\Delta^+$), $udd$ ($\Delta^0$), or $ddd$ ($\Delta^-$). The next
lowest-lying states come from the addition of one unit of orbital
angular momentum ($l_{\rho}=1$ or $l_{\lambda}=1$, and
$n_{\rho}=n_{\lambda}=0$). The states form the first band [$N = 2
(n_{\rho}+n_{\lambda}) +l_{\rho} + l_{\lambda} = 1$] of
negative-parity excited states produced via $L^P = 1^- \otimes
\{ S=\half$ or $S=\thalf \} \rightarrow J^P =
\{\half^-,\thalf^-,\fhalf^-\}$ to give the seven states
$2~N\half^-; \Delta\half^- ; 2~N\thalf^- ; \Delta\thalf^- ; N\fhalf^-$
which are constructed based on the overall symmetry of the combined
wave functions.

There are two main ingredients needed to complete such a
calculation. The first is a model of the spectrum and structure of
baryon states. The second is a model of baryon-baryon-meson vertices
and their momentum dependence.

The model of the spectrum must include not only states seen in
analyses of experimental data, but also states classified as
`missing', in order to provide wave functions for calculating the
vertices, and to know the thresholds associated with intermediate
states containing missing baryons. As mentioned above, the splittings
between states which result from differences in self energies can be
expected to be comparable to those that arise from the residual
interactions between the quarks. A complete calculation of the
spectrum therefore needs to adjust the residual interactions, and with
them the wave functions of the states used to calculate the
baryon-baryon-meson vertices, to account for these additional
splittings. The relativized quark potential model used in this work is
presented in more detail in Chapter~\ref{QuarkModel}, including the
changes required by the presence of the additional baryon-meson
intermediate states.

\begin{figure}
\begin{center}
\epsfig{file=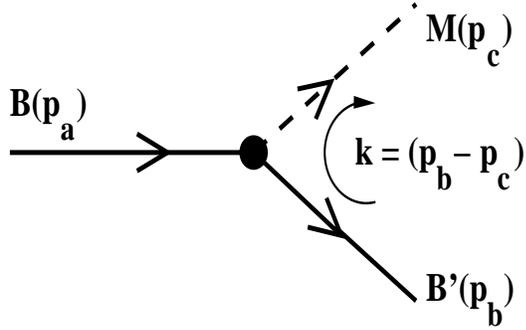,width=3in,height=2in,angle=0}
\caption[Momentum structure of the strong decay
vertex]{\label{relmom}Momentum structure of the strong decay vertex ${\cal
 M}_{BB'M}(k)$ in the initial baryon $B$ center-of-momentum frame.}
\end{center}
\end{figure}

The formalism used to model the strong decay vertices is presented in
more detail in Chapter~\ref{StrongDecays}. It provides an analytical
form of the momentum dependence of each vertex ${\cal
M}_{BB'M}(k)$ as a function of the relative momentum $k$
between the intermediate baryon $B'$ and meson $M$ in the
center-of-momentum frame of the initial baryon $B$ (see
Figure~\ref{relmom}).
Based on the lessons learned from earlier work, this calculation
includes all allowed combinations of intermediate states $B'M$ from
the sets
\begin{equation}
M \in \{ \pi,K,\eta,\eta',\rho,\omega,K^* \},\  B' \in \{
N,\Delta,\Lambda,\Sigma \}.
\end{equation}
including all excitations of the baryons states up to and including
the $N=3$ band states. Excited mesons states have been omitted at this
time as their higher mass and additional angular momentum highly
suppresses decays to these states. Studies including more massive
initial states should, however, consider including such states.

Based on the range of intermediate states included, it is worth noting
that for each $N^{(*)}$ initial state studied, the sum of
Eq.~\ref{loop_eq} ($\sum_{B'M}$) includes a total of 591 intermediate
baryon-meson states. Similarly, a total of 378 intermediate states are
included for each $\Delta^{(*)}$ initial state.

As the self energies due to a given intermediate state depend
crucially on the masses adopted for the intermediate hadrons, these
are taken to be the physical masses, where known, and model
masses~\cite{CaRo:1993} otherwise. The `bare' mass required to
reproduce the known physical mass of any initial baryon state $B$ considered
is found by solving the self-consistent (highly non-linear) equation 
\begin{equation}
\label{SE}
E + \Sigma_B(E) = M_B
\end{equation}
for the `bare' baryon mass $E_B^0$. Therefore the integration of
Eq.~\ref{loop_eq} needs to be performed over a range of bare masses
$E_B$ in order for the final result ($E_B^0$) to be extracted from the
intersection of the left side of Eq.~\ref{SE} with the right side when
$M_B=M_{\rm physical}$.
\chapter{The Quark Potential Model}
\label{QuarkModel}
The nonrelativistic constituent quark model (NRQM) owes its origin to
many authors but the model of Isgur and Karl and
collaborators~\cite{IsKa:1977,IsKa:1978a,IsKa:1978b,IsKa:1979} has
been qualitatively successful in both meson and baryon sectors,
despite its lack of theoretical foundation in QCD. In an effort to
correct some of the flaws of the NRQM, the Isgur-Karl model was later
`relativized' by Godfrey and Isgur~\cite{GoIs:1985}~(for mesons) and
Capstick and Isgur~\cite{CaIs:1986}~(for baryons). This latter version
has been used by Capstick and
Roberts~\cite{CaRo:1993,CaRo:1994,CaRo:1998} in extensive calculations
of strong decay amplitudes. A modified version of this model is used
in the present work to obtain the masses and wave functions of known
and `missing' baryon states. It is therefore appropriate to give an
overview of the main components and discuss the value of some of the
parameters used here.

The choice of dynamical degrees of freedom used to represent a baryon
depends on momentum transfer. At low $Q^2$, they can be taken to be
constituent quarks, which are valence quarks with effective masses of
about 220 MeV for $u$ and $d$ ($\sim$330 MeV in the NRQM), and about 420
MeV for the $s$ quark ($\sim$550 MeV in the NRQM). In this model the gluon
fields affect the quark dynamics by creating a confining potential in
which the quarks move. At short distances, a perturbative one-gluon
exchange between quarks is assumed to provide the spin-dependent
potential.
\section{The Hamiltonian}
The Hamiltonian used~\cite{CaIs:1986} for the baryon system has
the form
\begin{equation}
H = H_0 + V_{\rm oge} + V_{\rm conf}.
\end{equation}
where $H_0$ is the relativistic kinetic energy term
\begin{equation}
H_0 = \sum_{i=1}^3 (p_i^2 + m_i^2)^{1/2},
\end{equation}
$V_{\rm oge}$ is the one-gluon-exchange potential, and $V_{\rm conf}$ consists
of a string potential and its associated spin-orbit term arising from
the Thomas precession.

The one-gluon exchange potential has the form
\begin{equation}
V_{\rm oge} = \sum_{i<j} V_{ij}^{\rm oge}
\end{equation}
with the color induced interactions being
\begin{equation}
V_{ij}^{\rm oge} = V_{ij}^{\rm Coulomb} + V_{ij}^{\rm hyperfine} +
V_{ij}^{\rm spin-orbit(cm)}
\end{equation}
where 
\begin{equation}
V_{ij}^{\rm hyperfine} = V_{ij}^{\rm contact} + V_{ij}^{\rm tensor}.
\end{equation}
The Coulomb term is spin-independent and proportional to $1/r_{ij}$
(where ${\bf r}_{ij}$ is the relative position of the $(ij)$ quark pair),
the spin-orbit and hyperfine interactions are color-magnetic in nature, and the
hyperfine interaction consists of a Fermi contact term $\propto
\delta^3(r_{ij})$ and a tensor piece. The terms of the one-gluon
exchange potential can be found from the Breit-Fermi reduction of the
one-gluon exchange T-matrix element $\propto \bar{u}({\bf
p}',s')\gamma_{\mu}u({\bf p},s)$ where $u$ is approximated as the
Dirac four-spinor of a free particle.

The confining potential is composed of two parts
\begin{equation}
V_{\rm string} = b \sum_{i<j} r_{ij},
\end{equation}
and
\begin{equation}
V_{\rm spin-orbit(s)} = \sum_{i<j} V_{ij}^{\rm spin-orbit(Tp)}.
\end{equation}
The string part of $V_{\rm conf}$ is the adiabatic potential
corresponding to the energy of the minimum-length configuration of the
Y-shaped string linking the quarks. The spin-orbit term includes the
Thomas precession effects of the full spin-independent potential. For
ease of calculation, $V_{\rm string}$ is approximated by a sum of a
constant, an effective two-body piece, and a three-body piece
\begin{equation}
V_{\rm string} = C_{qqq} + fb\sum_{i<j}r_{ij} + V_{\rm 3b},
\end{equation}
where $C_{qqq}$ is an overall energy shift which arises from the
vacuum modifications due to the presence of colored fields in the
baryon, $f = 0.5493$ is chosen to minimize the size of the expectation
value of $V_{\rm 3b}$ in the harmonic oscillator ground state of the
baryon system, and $b$ is the meson string tension. The two-body part
of $V_{\rm string}$ is calculated directly during the diagonalization
of the Hamiltonian, and $V_{\rm 3b}$ is computed perturbatively. 

The potentials have been modified from their nonrelativistic
limit ($p/m \rightarrow 0$) by several effects. For example, since
constituent quarks are not point-like, the interquark coordinate is
smeared out over mass-dependent distances. The smearing is brought
about by convoluting the potentials with a function
\begin{equation}
\label{rho}
\rho_{ij}({\bf r}_{ij}) = \frac{\sigma_{ij}^3}{\pi^{\thalf}}
e^{-\sigma_{ij}^2 {\bf r}_{ij}^2}.
\end{equation}
where the $\sigma_{ij}$ are chosen to smear the interquark coordinate over
distances of approximately 0.22 fm for light quarks, and $O(1/M_Q)$ for
heavy quarks $Q$. A second modification allows the potentials to be momentum dependent
by introducing factors which replace quark mass terms by energy
dependent ones such as
\begin{eqnarray}
\beta_{ij} = 1 + \frac{p_{ij}^2}{(p_{ij}^2 + m_i^2)^{1/2}(p_{ij}^2 +
m_j^2)^{1/2}} \\
\delta_{ij} = \frac{m_im_j}{(p_{ij}^2 + m_i^2)^{1/2}(p_{ij}^2 +
m_j^2)^{1/2}}
\end{eqnarray}
where $p_{ij}$ is the magnitude of the momentum of either quark in
the $ij$ center-of-mass frame. These terms are included in the
potentials in the form of factors such as $(\delta_{ij})^{1/2 +
\epsilon_k}$ where the $\epsilon_k$'s are free parameters designed to
allow the rough description of the momentum dependence of each potential. 
\section{The Parameters}
For completeness, we include the final value of some of the
relativized quark potential model parameters used in this work. As
will be explained in more detail in Chapter~\ref{ResI}, the values
used in Ref.~\cite{CaIs:1986} and subsequent work were not adequate
here as they were meant to model a spectrum without taking into
consideration the existence of $qqq + q\bar{q}$
configurations. Therefore our work requires slightly different
parameters. Another reason for reducing the value of $\sigma_0$, which
corresponds to the inverse of a quark `size' and is one of the
smearing parameters used to define $\sigma_{ij}$ in Eq.~\ref{rho} (the
other parameter being $s$), is that it brings the electromagnetic form
factor of the quark required to fit nucleon electromagnetic form
factors in relativistic (light-cone based) models more in line with
this strong size. Studies have been done with various values of some
of the parameters to understand their effects before selecting the
final values. Some of the `intermediate' results will be presented in
Chapter~\ref{ResI} to illustrate this process. More information about
the different potentials, the origin and use of the parameters listed
here can be found in Ref.~\cite{CaIs:1986} and references within,
since their description is beyond the scope of this work.

It is important to note that all spin-orbit effects have been removed
from the Hamiltonian used to obtain the wave functions used in this
work. Studies on how best to introduce spin-orbit effects are in
progress but are secondary to the main goal of this work, and so are
not presented at this time.
\begin{table}
\caption[Relativized Quark Model Parameters.]{\label{Param}The
parameters of the relativized quark potential model.}
\vspace{0.5cm}
\begin{center}
\begin{tabular}{lccc}
\hline\hline
\multicolumn{1}{l}{Parameter}
& \multicolumn{1}{c}{This work}
& \multicolumn{1}{c}{Ref.~\cite{CaIs:1986}}
& \multicolumn{1}{c}{}\\
\hline
$\half(m_u + m_d)$ (MeV)  	& 220 	& Same	& Light quark mass \\[6pt]
$m_s$ (MeV)		      	& 419 	& Same  & Strange quark mass \\[6pt]
$b$ (GeV$^2$)			& 0.15	& Same  & String tension \\[6pt]
$\half + \epsilon_{\rm cont}$	& $\half - 0.168$  & Same &
Relativistic factor\\[6pt]
$\half + \epsilon_{\rm tens}$	& $\half - 0.168$  & Same &
``  \\[6pt]
$\half + \epsilon_{\rm Coul}$	& $\half$          & Same &
``  \\[6pt]
$\alpha_s^{\rm critical}$	& 0.550	& 0.60  & $\alpha_s(Q^2=0)$ \\[6pt]
$\sigma_0$(GeV)			& 0.833 & 1.80  & Relativistic smearing \\[6pt]
$s$				& 1.55	& Same  &  `` \\[6pt]
\hline \hline
\end{tabular}
\end{center}
\end{table}
\chapter{Modeling Strong Decays}
\label{StrongDecays}
\section{Introduction}
One very important element of our calculation is a model of the
momentum dependence of each vertex in Figure~\ref{mixing}~(a). If that
diagram were to be cut in half, one could see that each half
represents a decay $B \rightarrow B'M$. We can therefore use a decay
model to obtain the structure of the vertex and hence its momentum
dependence. The two most suitable classes of models for this work are
briefly described below before going into the details of the specific
model used here (for a recent review of these and other strong decay
models see Ref.~\cite{CaRo:2000}).

The first class, known as elementary-meson-emission (EME) models, has
baryons treated as objects with a quark structure while mesons are
treated as elementary, point-like objects emitted from a quark during
the decay. Each decay transition is described in terms of a coupling
constant. This implies many parameters, although several coupling constants
can be approximately related via SU(2) or SU(3) flavor symmetry. This
class of models lends itself well to relativistic treatment, which is
often desirable for light mesons such as the pion. Unfortunately since
mesons are modeled as point-like objects, treatment of excited mesons
is restricted since radial excitations imply an extended spatial wave
function which is not modeled.

The other class of models, referred to broadly as pair creation
models, treat all hadrons as composite objects. The decay of a hadron
coincides with the creation of a quark-antiquark pair somewhere in the
hadronic medium. The created antiquark then combines with a quark of
the original hadron to create a daughter meson while the created quark
becomes part of the other daughter hadron. These models describe
hadron emission in a unified way, often involving only one free
parameter (the pair-creation strength $\gamma$), and allow for the
treatment of all excited baryons and mesons within the same
framework. In contrast to EME models, pair creation models are
non-relativistic and therefore approximations are made. They are
nonetheless more realistic, simple, and have been successfully applied
to the study of a broad range of strong decays for both mesons and
baryons.

There are several types of pair creation models, such as the $\tp0$
and $^3S_1$ models (named after the quantum numbers of the created
pair), and the flux-tube and string breaking models, where the
location of the created pair is restricted to an area inside the
chromoelectric flux tube (the tube of gauge field that is shown by
lattice QCD to form between two colored sources) or along the string
axis. We describe our choice in some detail in the next section.
\section{The $\tp0$ Model}
Due to its simplicity and past successful applications to the strong
decays of hadrons, the $\tp0$ model, popularized by Le Yaouanc {\it et
al.}~\cite{LYaOPR} has been selected to be used in this research. It
has been widely applied to baryon
decays~\cite{CaRo:1993}~\cite{CaRo:1994}~\cite{CaRo:1998}, meson
decays, and even generalized to the decay of states composed of
n-quarks~\cite{RoSB:1992}.

Within this model, the strong decay can be seen as a process where a
quark-antiquark pair is created from the QCD vacuum with quantum
numbers $J^{PC}= 0^{++}$. As shown below, in the $^{2S+1}L_J$
notation, this corresponds to $\tp0$, hence the name of the model. The
pair can be created anywhere in space, but wave function overlaps will
naturally strongly suppress creation very far from the initial
hadron. The created pair is added to the initial system, giving rise
to two new non-interacting final state hadrons. To be observed, these
new hadrons must be color singlets. Additionally, the $q\bar{q}$ pair
must be neutral with respect to the additive quantum numbers, meaning
that it must also be a flavor singlet, and have zero total angular
momentum. Because the quark and antiquark have opposite intrinsic
parity, parity conservation further dictates the pair be in a relative
p-wave (i.e. $L=1$) so that its total spin must be one ($S=1$) to
combine to the required $J=0$.

\begin{figure}
\begin{center}
\epsfig{file=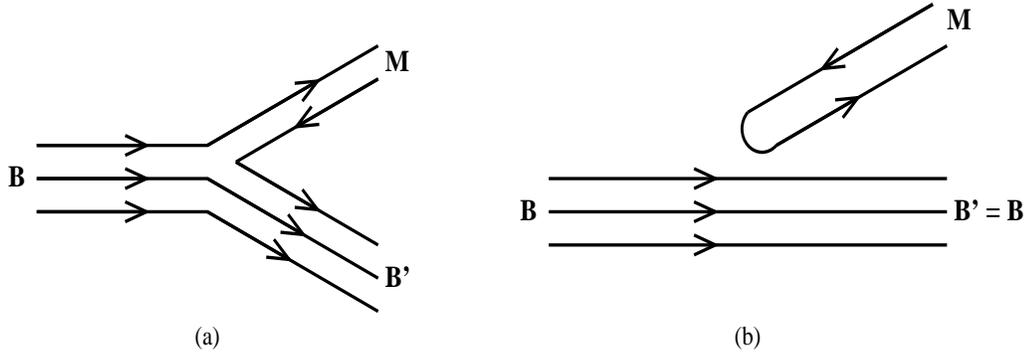,width=5.5in,height=2in,angle=0}
\caption[OZI processes.]{\label{ozi}OZI-allowed (a) and suppressed
(b) processes $B \rightarrow B'M$, in a quark pair creation model.}
\end{center}
\end{figure}

It is important to note that only Okubo-Zweig-Iizuka (OZI) allowed
decays are considered, as illustrated in Figure~\ref{ozi}(a). A
process is said to be OZI-forbidden, or suppressed, [see
Figure~\ref{ozi}(b)] if the quark of the created quark pair does not
combine with quarks in the initial hadron but instead the created
quark and antiquark form a separate meson.
%
\subsection{The Operator}
The starting point in modeling the $B \rightarrow B' M$ transitions of
baryons in the $\tp0$ model is the form of the operator $T$
responsible for the decay. Within this model, the operator does
not result from a detailed Hamiltonian that would come from the QCD
Lagrangian, as the complexity would be overwhelming. Instead it is
entirely phenomenological and is defined only for the decay process
under consideration. It has the following form
\newpage
\begin{eqnarray} \label{3P0op}
T&=&-3\gamma\sum_{i,j} \int d {\bf p}_i d {\bf p}_j \  \delta( {\bf p}_i +
{\bf p}_j)\  C_{ij} \  F_{ij}\  e^{-f^2({\bf p}_i - {\bf p}_j)^2}\nonumber\\
&\times& \sum_m \langle 1,m;1,-m|0,0 \rangle \  \chi_{ij}^m \  { \cal
Y}_1^{-m}( {\bf p}_i - {\bf p}_j ) \ b_i^{\dagger}( {\bf p}_i ) \  d_j^{\dagger}( {\bf p}_j ),
\end{eqnarray}
where $C_{ij}$ and $F_{ij}$ are the color and flavor wave functions of
the created pair, both assumed to be singlet, $\chi_{ij}$ is the spin
triplet wave function of the pair, and ${\cal Y}_1({\bf p}_i-{\bf
p}_j) = | {\bf p}_i-{\bf p}_j | \ Y_1(\widehat{{\bf p}_i-{\bf p}_j})$
is the solid harmonic indicating that the pair is in a relative p-wave
$(L=1)$. Note that the threshold behavior resulting from this $|{\bf
p}_i-{\bf p}_j|$ factor is as seen experimentally. Here
$b_i^{\dagger}( {\bf p}_i\/ )$ and $d_j^{\dagger}( {\bf p}_j\/)$ are
the creation operators for a quark and an antiquark with momenta ${\bf
p}_i$ and ${\bf p}_j$ respectively. The exponential has been
introduced to give the vertex a spatial extent by creating the
quark-antiquark pair over a smeared region, instead of at a point as
is the case in the usual version of the $\tp0$ model. The addition of
this form factor `softens' the vertices and suppresses the self energy
contributions from intermediate states where the hadrons have high
relative momentum.

There are only two phenomenological parameters in this model. The
first one is $\gamma$, the coupling strength, which we fit to the
experimentally well known $\Delta \rightarrow N \pi$ decay, and the
second one is $f$, which is set to give a reasonable
quark-pair-creation vertex size of around 0.35 fm (the same as that
used in Geiger and Isgur~\cite{GeIs:1991-2}~and Silvestre-Brac and Gignoux~\cite{SBGi:1991}).

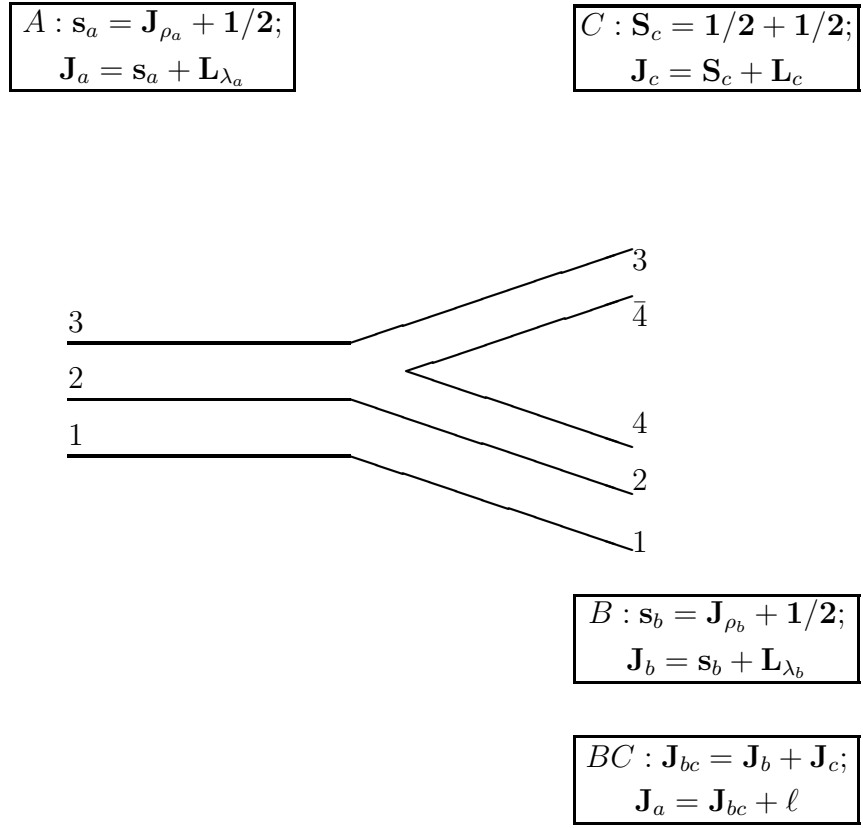
\begin{figure}
\thicklines 
\unitlength 0.75cm
\begin{center}
\begin{picture}(16,18)
\multiput(2.0,8.0)(0.0,1.0){3}{\line(1,0){5}} 
\multiput(7.0,8.0)(0.0,1.0){2}{\line(3,-1){5.0}} 
\put(2.0,8.2){$1$}
\put(2.0,9.2){$2$}
\put(2.0,10.2){$3$}
\put(12.0,11.3){$3$}
\put(12.0,10.3){$\bar 4$}
\put(12.0,6.3){$1$}
\put(12.0,7.4){$2$}
\put(12.0,8.4){$4$}
\put(7.0,10.0){\line(3,1){5.0}} 
\put(8.0,9.5){\line(3,1){4.0}} 
\put(8.0,9.5){\line(3,-1){4.0}} 
\put(1.0,14.5){\framebox(5.0,1.5){\shortstack{$A:{\bf s}_a={\bf J}_{\rho_a}+{\bf 
1/2};$\\ {}\\ ${\bf J}_a={\bf s}_a+{\bf L}_{\lambda_a}$}}} 
\put(11.0,4.0){\framebox(5.0,1.5){\shortstack{$B:{\bf s}_b={\bf J}_{\rho_b}+{\bf 
1/2};$\\ {}\\ ${\bf J}_b={\bf s}_b+{\bf L}_{\lambda_b}$}}} 
\put(11.0,14.5){\framebox(5.0,1.46){\shortstack{$C:{\bf S}_c={\bf 1/2}+{\bf 
1/2};$\\ {}\\ ${\bf J}_c={\bf S}_c+{\bf L}_c$}}} 
\put(11.0,1.5){\framebox(5.0,1.5){\shortstack{$BC:{\bf J}_{bc}={\bf J}_b + 
{\bf J}_c;$\\ {}\\ ${\bf J}_a={\bf J}_{bc}+{\bf \ell}$}}} 
\end{picture} 
\end{center}
\caption[Schematic diagram of the decay]{\label{decay}Schematic diagram of the
decay $B\rightarrow B'M$ in the $^3P_0$ model. The angular momentum
notation is shown. The decay proceeds through $B(123)\rightarrow
12(4\bar{4})3\rightarrow B'(124)M(\bar{4}3)$.}
\end{figure}

Consider an initial observable system $A$ (a baryon composed of three
quarks) decaying into two observable, non-interacting hadrons; baryon
$B$ and meson $C$. One quark from $A$ will merge with the created
antiquark to form meson $C$, and the remaining two `initial' quarks
will merge with the created quark to form baryon $B$. The notation
used is illustrated in Figure~\ref{decay}. Note that in this version
of the $\tp0$ model, quarks 1 and 2 are considered spectators as they
do not participate in the decay.
\subsection{Wave Functions Considerations}
For the transition $A\rightarrow BC$, we are interested in evaluating
the following transition amplitude
\begin{equation}\label{amp1}
M= \langle BC|T|A \rangle,
\end{equation}
where $|A\rangle$ denotes the wave function of the initial baryon $A$,
and $|BC\rangle$ the wave function of the final baryon-meson pair. The
initial system is assumed to be in a static state (made up of a quantum
superposition of harmonic oscillator substates), multiplied by a plane
wave with relative state momentum ${\bf K}_a$ for center-of-mass
motion, where $a$ denotes the quantum numbers needed to describe the
basis states. The total wave function for the initial state
$|A\rangle$ expressed in momentum representation is expanded in terms
of basis states $\Psi_a^A$
\begin{equation}\label{PsiA}
\Psi^{A,{\bf K}_a} = \delta({\bf p}_a - {\bf K}_a) \Psi^A = \sum_{a}
{d^A}_a \Psi_a^{A,{\bf K}_a}.
\end{equation}

The coefficients ${d^A}_a$ are obtained by diagonalization of the
Hamiltonian $H$ in the basis of the $\Psi_a^A$, taken here to be
harmonic-oscillator basis states. The series is truncated to
$N=2(n_{\rho}+n_{\lambda})+l_{\rho}+l_{\lambda}=6$ for the
positive-parity states and $N=7$ for negative-parity states, giving of
the order of 100 substates for each $J^P$. Note that the level $N$ of
the expansion is related to the sum of the powers of the two
coordinates $\lpmb{\rho}$ and $\lpmb{\lambda}$ (see
Appendix~\ref{ap:WF}) in the associated Laguerre polynomials. A higher
power means a shorter length scale, therefore the maximum $N$ was
chosen to `resolve' the shortest range interaction in the Hamiltonian,
or equivalently allow the variational calculation of the energy to
converge. The expansion coefficients $d_a^A$ are such that the total
wave function $\Psi^{A,{\bf K}_a}$ is antisymmetric, despite the fact
that the basis states $\Psi_a^{{\bf K}_a}$ are taken to be
antisymmetric only in the first two quarks.

The wave functions for the final baryon and meson are given in a
similar fashion
\begin{equation}\label{PsiB}
\Psi^{B,{\bf K}_b} = \delta({\bf p}_b - {\bf K}_b) \Psi^B = \sum_{b}
d_b^B {\Psi_b^{B,{\bf K}_b}}.
\end{equation}
\begin{equation}\label{PsiC}
\Psi^{C,{\bf K}_c} = \delta({\bf p}_c - {\bf K}_c) \Psi^C = \sum_{c}
d_c^C \Psi_c^{C,{\bf K}_c}.
\end{equation}
By combining $\Psi^{B,{\bf K}_b}$ and $\Psi^{C,{\bf K}_c}$ we obtain a
wave function $|BC;J_b M_b {\bf K}_b J_c M_c {\bf K}_c \rangle$ which
describes the hadrons in a plane wave with their angular momenta
decoupled. For ease in further treatment of angular momenta, we couple
${\bf J}_b + {\bf J}_c = {\bf J}_{bc}$, and change the variables ${\bf
p}_b$ and ${\bf p}_c$ to ${\bf K} = {\bf p}_b + {\bf p}_c$ and ${\bf
k} = \half({\bf p}_b - {\bf p}_c)$. Then ${\bf K}$ represents the
total momentum of the $BC$ system and is conserved through the term
$\delta({\bf p}_a - {\bf K}_a)$ of equation~\ref{PsiA}, and ${\bf k}$
is the relative momentum between $B$ and $C$. Instead of using a plane
wave $|{\bf k}\rangle$, we change to a spherical wave $|lmk\rangle$ via 
\begin{equation}
\langle{\bf p}_b{\bf p}_c|{\bf K}_0 lmk_0\rangle = \delta({\bf K} -
{\bf K}_0) \frac{{Y^m}_l(\bf \hat{k})}{k^2}\delta(k - k_0).
\end{equation} 
Finally, the relative momentum $l$ is coupled to $J_{bc}$ to give the
total angular momentum $J_a$, so that the form of the final state
wave function becomes
%
\begin{eqnarray}
\lefteqn{|BC,J_bJ_c,J_{bc}l;J_aM_a;{\bf K}_0k_0\rangle} \nonumber \\
& = & \int d{\bf K}_b d{\bf K}_c \sum_{M_{bc},m,M_b,M_c}
\langle J_bM_bJ_cM_c|J_{bc}M_{bc}\rangle \nonumber \\
& & \times \langle J_{bc}M_{bc}lm|J_aM_a\rangle \langle {\bf K}_b{\bf
K}_c|{\bf K}_0lmk_0\rangle |BC,J_bM_b{\bf K}_bJ_cM_c{\bf K}_c\rangle.
\end{eqnarray}

Baryon states are written as
\begin{equation}
\Psi = C_A \phi \sum \psi \chi.
\end{equation}
where $C_A$, $\phi$, $\psi$, and $\chi$ are the color, flavor,
spatial, and spin wave functions respectively. The baryon wave functions used
in this calculation were produced using a relativized
model~\cite{CaIs:1986} with variable-strength spin-dependent
(one-gluon exchange) contact, tensor, and spin-orbit interactions
between the quarks. More details about the baryon wave functions can be
found in Appendix~\ref{ap:WF}.
\subsection{Transition Amplitude}
From equation~\ref{3P0op}, the transition amplitude is not Galilean
invariant since it contains a factor $\delta {\bf p}$, where ${\bf p}$
is evaluated in a definite frame. The results therefore depend on the
chosen frame of reference. A good choice of frame is the one
where the decaying baryon $A$ is at rest, so we set ${\bf K}_a =
0$. Momentum conservation yields a factor $\delta({\bf K}_0)$ in the
amplitude, and we now rewrite equation~\ref{amp1} as
\begin{equation}
\langle BC|T|A \rangle = \delta({\bf K}_0)M_{A \rightarrow BC}.
\end{equation}
Incorporating equations~\ref{PsiA},~\ref{PsiB}, and~\ref{PsiC}, we obtain the
expression
\begin{equation}
M_{A\rightarrow BC} = \sum_{a,b,c} d_b^{B*} d_c^{C*} d_a^A
M_{A\rightarrow BC}(a,b,c)
\end{equation}
The color, flavor, spin, and spatial degrees of freedom can be
separated via invariance techniques. The resulting amplitude then
reduces to products of sums over internal summation variables of $6-j$
and $9-j$ coefficients from angular momentum recoupling, and flavor,
and spatial matrix elements that can be calculated independently,
which then can be combined to give a total decay matrix element.  The
final form of $M_{A\rightarrow BC}(a,b,c)$ is
%
\begin{eqnarray}
\label{ampl}
&&M_{A \to BC}(a,b,c)={6\gamma \over 3 \sqrt{3}}(-1)^{J_a+J_b+\ell_a+\ell_b
-1}\sum_{J_\rho,s_a,s_b}
 {\hat J}\/_\rho^2 \hat s_a {\hat S}\/_a {\hat L}\/_a \hat s_b {\hat S}\/_b 
{\hat L}\/_b\nonumber \\
&&\left \{\matrix{S_a&L_\rho&s_a\cr
               \ell_a&J_a&L_a\cr} \right \}
\left\{\matrix{L_\rho&S_\rho&J_\rho\cr
                {1 \over 2}&s_a&S_a} \right\} 
\left \{\matrix{S_b&L_\rho&s_b\cr
               \ell_b&J_b&L_b\cr} \right \}
\left\{\matrix{L_\rho&S_\rho&J_\rho\cr
                {1 \over 2}&s_b&S_b} \right\} \nonumber \\
&& (-1)^{\ell+\ell_a+J_c-L_c-S_c}{\cal F}(ABC) \nonumber \\
&&\times \sum_{S_{bc}}(-1)^{s_a-S_{bc}} 
\left [\matrix{J_\rho& \half &s_b\cr
               \half & \half &S_c\cr
               s_a&1&S_{bc}} \right ]
\sum_{L_{bc}} (-1)^{L_{bc}} 
               \left [\matrix{s_b&\ell_b&J_b\cr
                              S_c&L_c&J_c\cr
                              S_{bc}&L_{bc}&J_{bc}} \right ]\nonumber \\
&&\times \sum_L {\hat L}\/^2
 \left\{\matrix{s_a&\ell_a&J_a\cr
                L&S_{bc}&1} \right\}
\left\{\matrix{S_{bc}&L_{bc}&J_{bc}\cr
                 \ell&J_a&L} \right\} \varepsilon(\ell_b,L_c,L_{bc},\ell,\ell_a,L,k_0),
\end{eqnarray}
where the factor of $6$ comes from the redefinition, for the created
pair, ${\bf P}={\bf p}_i + {\bf p}_j$ and ${\bf p} = \half ({\bf p}_j
- {\bf p}_i)$ so that the spherical harmonic found in the operator (eq.~\ref{3P0op}) can be rewritten as $-3\gamma
{\cal Y}_1(-2{\bf p}) = 6 \gamma {\cal Y}_1({\bf p})$. The overall factor of
$\frac{1}{3}$ is the color matrix element, ${\cal F}$ is the flavor
overlap, and $\varepsilon$ is the spatial matrix element. 

Further explanation of notation and derivations of some components have
been gathered in appendix~\ref{ap:TA} for the interested reader. 
\chapter{Results}
\label{ResI}
Graphical and tabulated results for the self energies of the ground
state Nucleon, $\Delta$ and non-strange $L=1$ negative-parity states
are presented in the following sections. First, evidence of the
convergence of our results is shown, indicating that a minimum number
of intermediate states have been included in order to obtain stable
and reliable results. The same graphs also show the effects of the
various decay thresholds, and their effect on the sums of self
energies. Next, tables displaying the impact of changes in the
Hamiltonian and the associated baryon model wave functions are
presented. Finally the $qqq$ spectrum obtained from the modified
Hamiltonian is graphically compared with the spectrum of bare energies
obtained from fitting the sum of the bare energies and self energies
to the physical masses. This illustrates that it is possible, in a
self-consistent calculation, to describe the observed masses with a
combination of splittings induced by interquark forces and differences
in the self energies.
\section{Convergence and Thresholds}
One important result coming out of this calculation is the phenomenon of
convergence. As pointed out before, the number and type of
intermediate states included in this type of calculation can
dramatically change the final results. In the figures that follow,
this concept and the associated consequences will be illustrated for the
states studied.

Figures~\ref{Nuc}~through~\ref{N5m1} show the self-energy
contributions to the masses of several baryons, from the sum of
intermediate $B'M$ states for $B'$ including progressively higher
harmonic oscillator bands, and $M$ the complete set of pseudo-scalar
and vector mesons ({\it
i.e.}~$\pi,K,\eta,\eta',\rho,\omega,K^*$). These figures show the
results obtained from using wave functions created from a Hamiltonian
that includes both the contact and the tensor part of the hyperfine
interaction. In each figure, each line represents the sum $E +
\Sigma_{B}(E)$ with $\Sigma_{B}(E)$ obtained with a set of baryons
$B'$ in the harmonic oscillator bands indicated in the subscript
[$\Sigma(E)_{N=0}$, $\Sigma(E)_{N\leq 1}$, $\Sigma(E)_{N \leq 2}$, and
$\Sigma(E)_{N\leq 3}$]. For each of these sums of intermediate states,
the corresponding bare mass $E_{B}^0$ can be extracted by reading the
value of the energy $E$ corresponding to the intersection between the
curve for the sum of self energies $E+\Sigma_{B}(E)$ and the
horizontal line representing the physical mass. This process is
in effect solving
\begin{equation}
\label{selfe}
E + \Sigma_{B}(E) = M_B,
\end{equation}
for $E=E_B^0$ when $M_B=M_{\rm physical}$ with progressively larger sum over bands of baryon
intermediate states $N=0$, $N\leq 1$, $N\leq 2$, and $N\leq 3$.

As will be seen in some of the figures and tables that follow,
occasionally more than one solution is possible for Eq.~\ref{selfe}
due to oscillations caused by the presence of $B'M$ decay thresholds
(the energies at which the decays $B\rightarrow B'M$ become physical, {\it
i.e.}~energetically possible). In these few cases a range of values is
presented unless it is clear that one solution is favored. More
details are presented in the next section.
\subsection{Nucleon and Delta Ground States}
\begin{figure}
\begin{center}
\epsfig{file=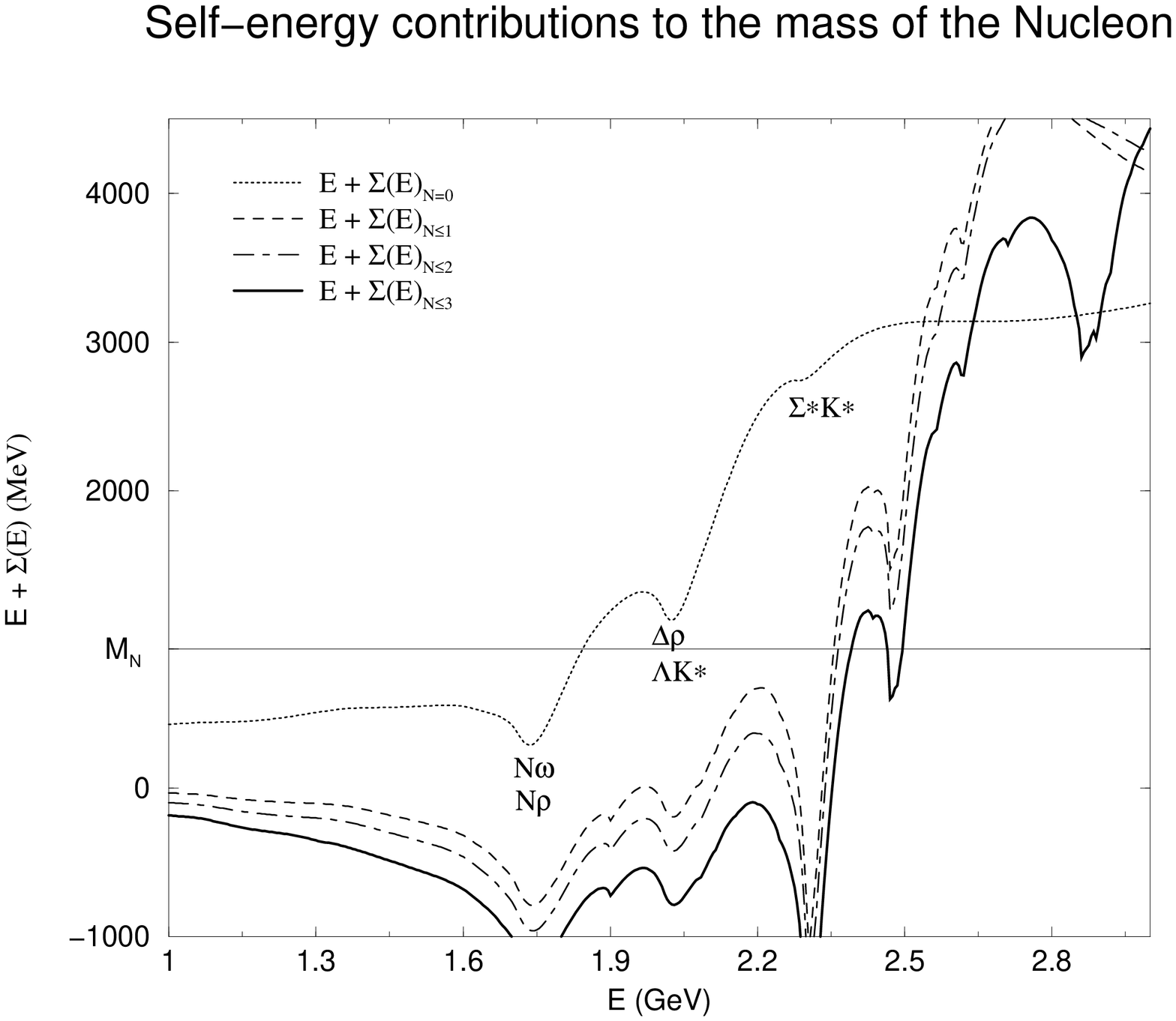,width=5.5in,height=6in,angle=270}
\caption[Self-energy contributions to the mass of the
Nucleon]{\label{Nuc}Sum of the bare energy and self energies as a
function of the bare energy for the Nucleon ground state with
$\alpha_s=0.55$ and $\alpha=0.5$~GeV.}
\end{center}
\end{figure}

Figure~\ref{Nuc} and Table~\ref{halfat_4}(b) show the self energy
contributions to the mass of the Nucleon from a progressively larger
sum of intermediate baryon-meson states. The first point to notice is
the sizeable difference between the `bare mass' $E_B^0$ when only
baryons in the $N=0$ band are included, {\it i.e.}~$E_B^0=1.85$~GeV, and
when other baryon states are included, $E_B^0=2.36$ to $2.50$~GeV. These
bare masses are not observables, but the mass splitting between the
nucleon and other states is, therefore any large variation in the bare
mass of the nucleon can change its relationship with other states in
the spectrum.

Next note that this `bare mass' difference is not `built' of equal
contributions from each set of intermediate states but comes mostly
from the inclusion of the $N=0$ and $N=1$ states (first band of
negative-parity excited states), indicating that the ground state
nucleon couples more strongly to states in these bands. Here the $N=0$
line can be thought of as a starting point for this work, since the
best of the previous studies of this state include only the
intermediate states included in this band, or restricted the
intermediate mesons to only the pion. The addition of the $N=2$ and
$N=3$ bands of states changes the bare mass by a very small amount
indicating that the sum over intermediate states has converged and a
stable solution has been reached for the ground state nucleon. This is
a very important result, and it will be shown that a lack of
convergence can greatly affect the final results of such
calculations. As will be seen below, the impact of the different bands
of states varies with the initial state studied, and the inclusion of
the $N=2$ and $N=3$ band baryons is important for other states. In the
case of the nucleon, these bands were added for consistency.

An additional item that needs explanation is the presence of multiple
solutions for Eq.~\ref{selfe} when all intermediate baryon states up
to the $N=3$ harmonic oscillator band are included (solid line in
Figure~\ref{Nuc}). This stems from the presence of decay thresholds
and their effect on the size of the self energies. The locations of
some ground state thresholds are labeled on the figure, but others
cannot be identified due to the large number of $B'M$ states
included. It is possible that these threshold effects could be
`dampened' by the inclusion of the widths of intermediate particles as
an imaginary part in the energy denominator of Eq.~\ref{loop_eq}, but
that is a higher-order effect which remains to be investigated. In
selecting a favored solution for the nucleon, studies of the
dependence of its self energy on the wave functions have shown that
the second and third solutions sometimes vanish, but the first
solution is always present. Therefore it seems prudent to retain only
the lower bare mass for the nucleon until more consistent results are
obtained for the other masses.

\begin{figure}
\begin{center}
\epsfig{file=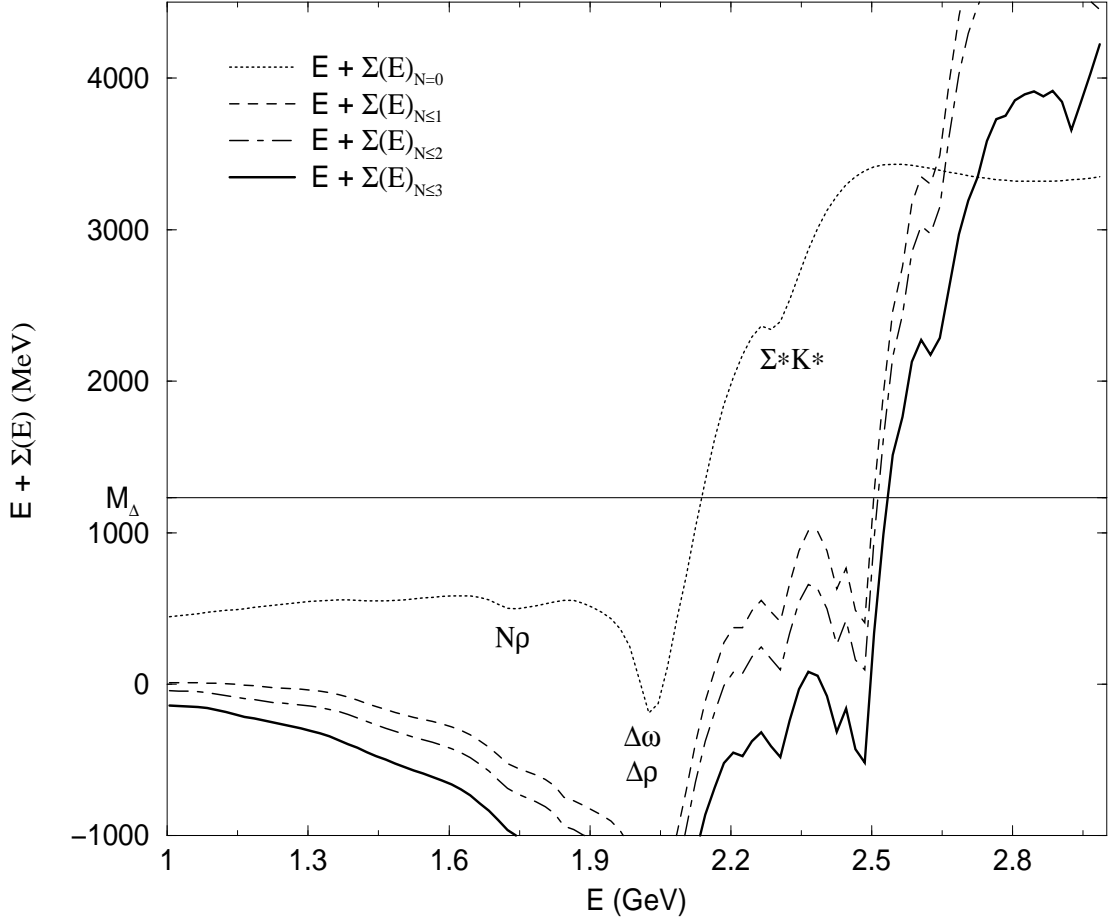,width=5.5in,height=6in,angle=270}
\caption[Self-energy contributions to the mass of the $\Delta$]{\label{Del}Sum of the bare energy and self energies as a
function of the bare energy for the $\Delta$ ground state with
$\alpha_s=0.55$ and $\alpha=0.5$~GeV.}
\end{center}
\end{figure}

Figure~\ref{Del} and Table~\ref{halfat_4}(b) show similar results for
the ground state $\Delta$. Again the evolution of the bare masses is
clear and once more the $N=0$ and $N=1$ band baryon states are seen to
make the largest contributions to the self energies. Here the
convergence is even more apparent, as the addition of the $N=2$ and
$N=3$ band states only resulted in an overall downward shift of the
$E+\Sigma_B(E)$ curves with very little movement along the energy
axis. The almost vertical slope of the last three lines clearly
indicates that not only has the sum over intermediate states
converged, but that the inclusion of additional states would have an
insignificant impact on the final bare mass.

Note that the importance of the labeled thresholds in the case of the
$\Delta$ is different than those for the nucleon, revealing some of
the differences in the internal structure of these two states. This
reflects the results of similar strong decay calculations which
predict where to look experimentally for resonances by highlighting
strong coupling to certain decay channels over others. These
calculations also explain why some resonances remain `missing'; they
couple weakly to experimentally accessible decay channels.

Combination of the results of Figures~\ref{Nuc} and~\ref{Del} [also
see Table~\ref{halfat_4}(b)] shows that if only intermediate ground
state baryons ($N=0$ band) are included, the $N$-$\Delta$ splitting is
roughly $290$~MeV. When states in the $N=1$ band are included the
splitting is reduced to about $140$ MeV, and that result remains
mostly unchanged by the addition of baryon intermediate states in the
$N=2$ and $N=3$ bands. This agrees well with the expectation from
other models (see Ref.~\cite{BHF:1996} where the authors find that
within their model, 2/3 of the $N$-$\Delta$ mass splitting comes from
one-gluon exchange effects, with the remaining third coming from
pion-exchange) that a substantial portion of the $N$-$\Delta$
splitting should come from a source other than the quark-quark
residual interactions, in this case a difference in self energies due
to all $B^\prime M$ intermediate states. This result will be shown to
hold despite changes to the wave functions from variations in the
quark residual interactions.
\subsection{Non-Strange $L=1$ Negative-Parity Baryons}
\begin{figure}
\begin{center}
\epsfig{file=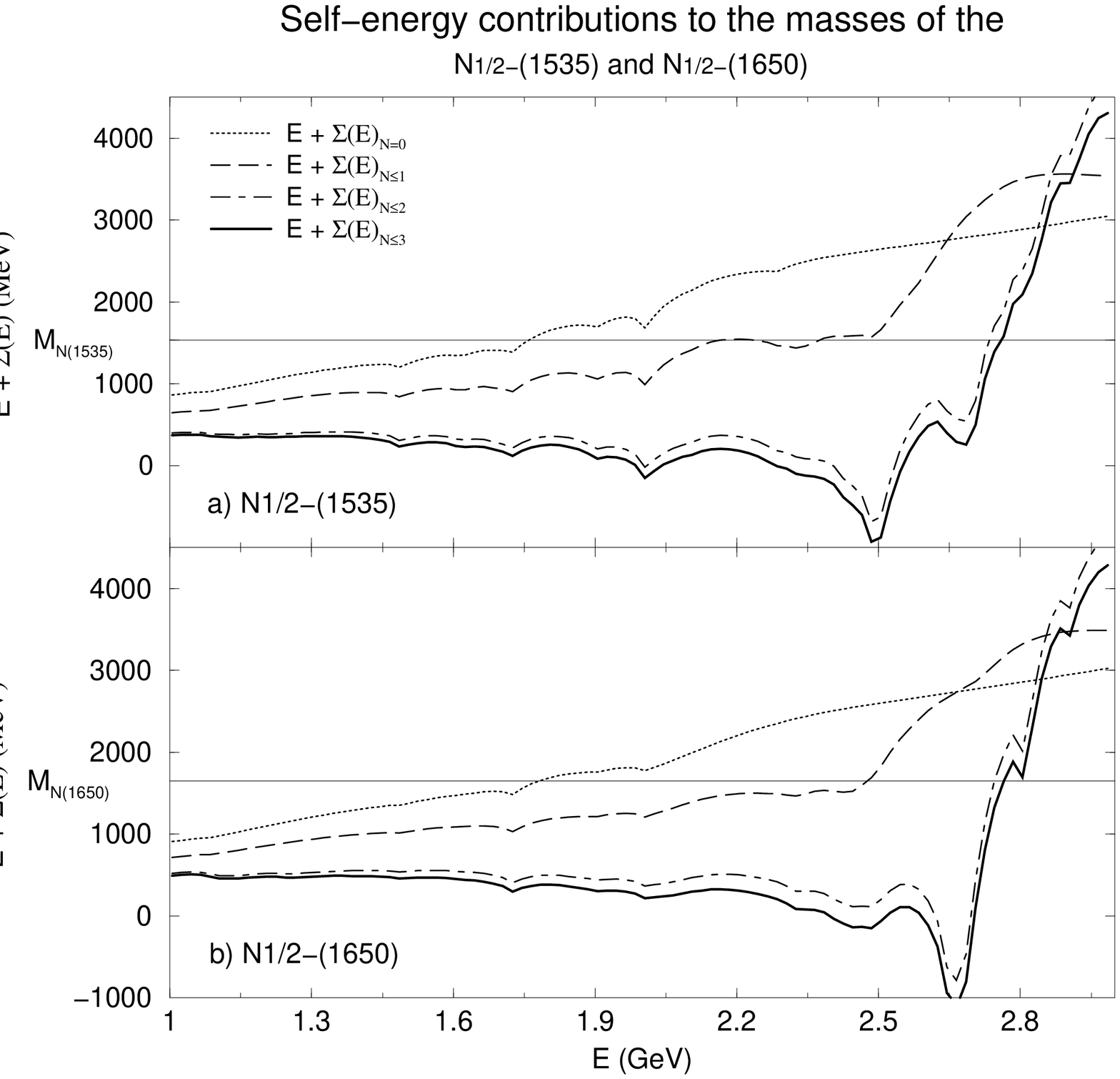,width=7in,height=6in,angle=270}
\caption[Self-energy contributions to the masses of $N\half^-(1535)$
and $N\half^-(1650)$]{\label{N1m1_N1m2}Sum of the bare energy and self
energies as a function of the bare energy for a) $N\half^-(1535)$ and
b) $N\half^-(1650)$ with $\alpha_s=0.55$ and $\alpha=0.5$~GeV.}
\end{center}
\end{figure}

\begin{figure}
\begin{center}
\epsfig{file=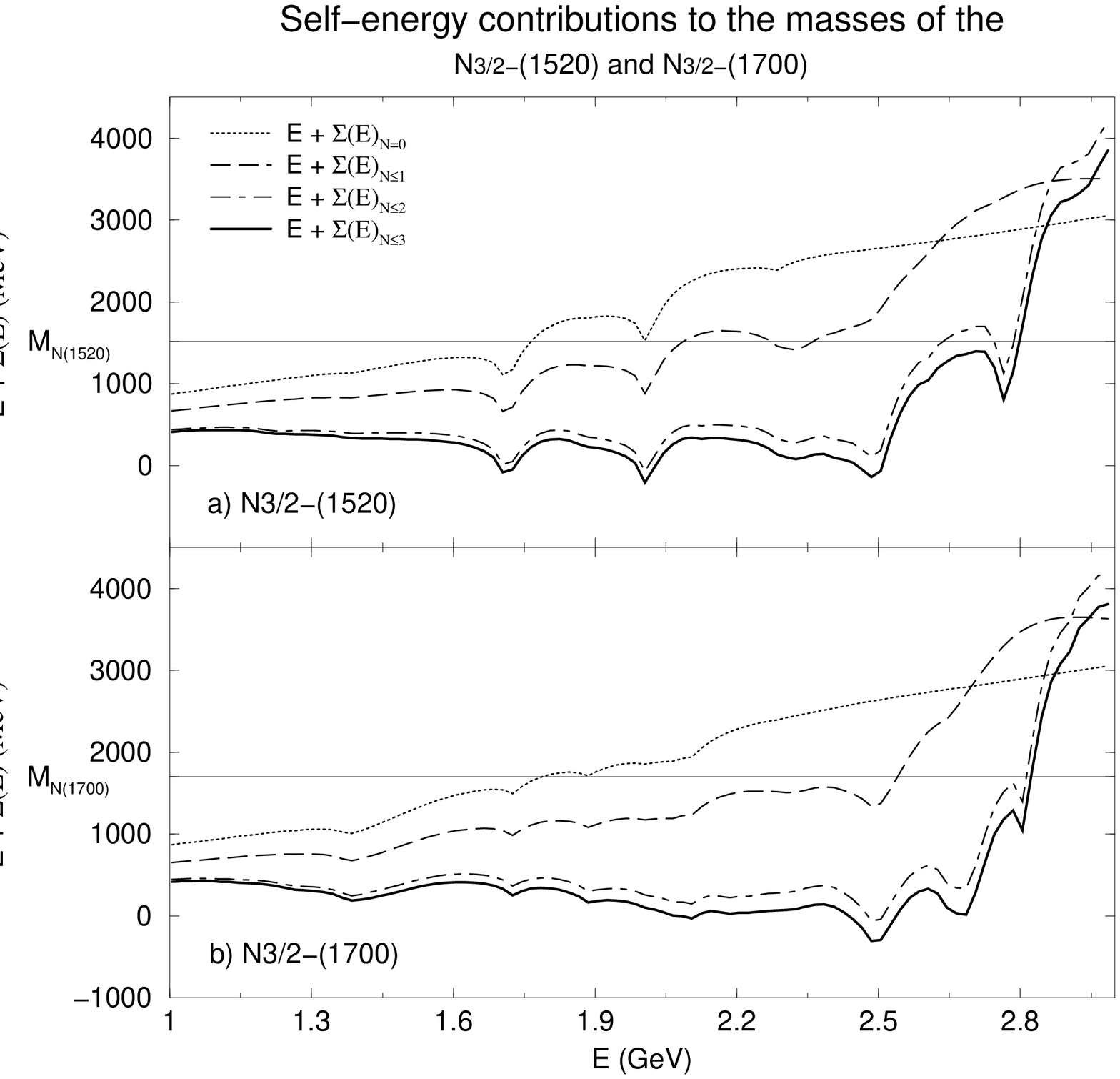,width=7in,height=6in,angle=270}
\caption[Self-energy contributions to the masses of
$N\thalf^-(1520)$ and $N\thalf^-(1700)$]{\label{N3m1_N3m2}Sum of the
 bare energy and self energies as a function of the bare energy for a)
 $N\thalf^-(1520)$ and b) $N\thalf^-(1700)$ with $\alpha_s=0.55$ and
 $\alpha=0.5$~GeV.}
\end{center}
\end{figure}

\begin{figure}
\begin{center}
\epsfig{file=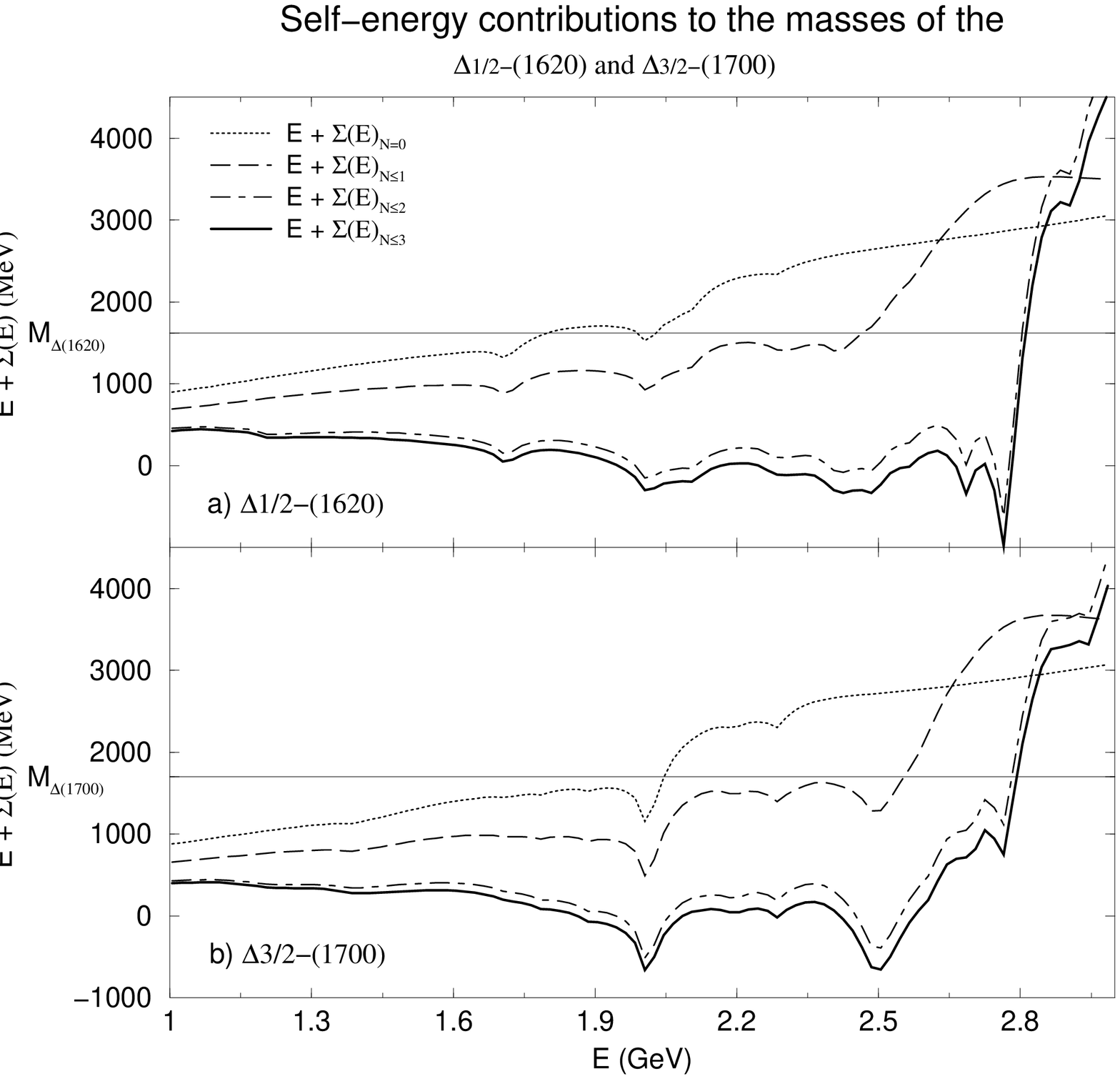,width=7in,height=6in,angle=270}
\caption[Self-energy contributions to the masses of $\Delta\half^-(1620)$ and $\Delta\thalf^-(1700)$]{\label{D1m1_D3m1}Sum of the
 bare energy and self energies as a function of the bare energy for a)
 $\Delta\half^-(1620)$ and b) $\Delta\thalf^-(1700)$ with
 $\alpha_s=0.55$ and $\alpha=0.5$~GeV.}
\end{center}
\end{figure}

\begin{figure}
\begin{center}
\epsfig{file=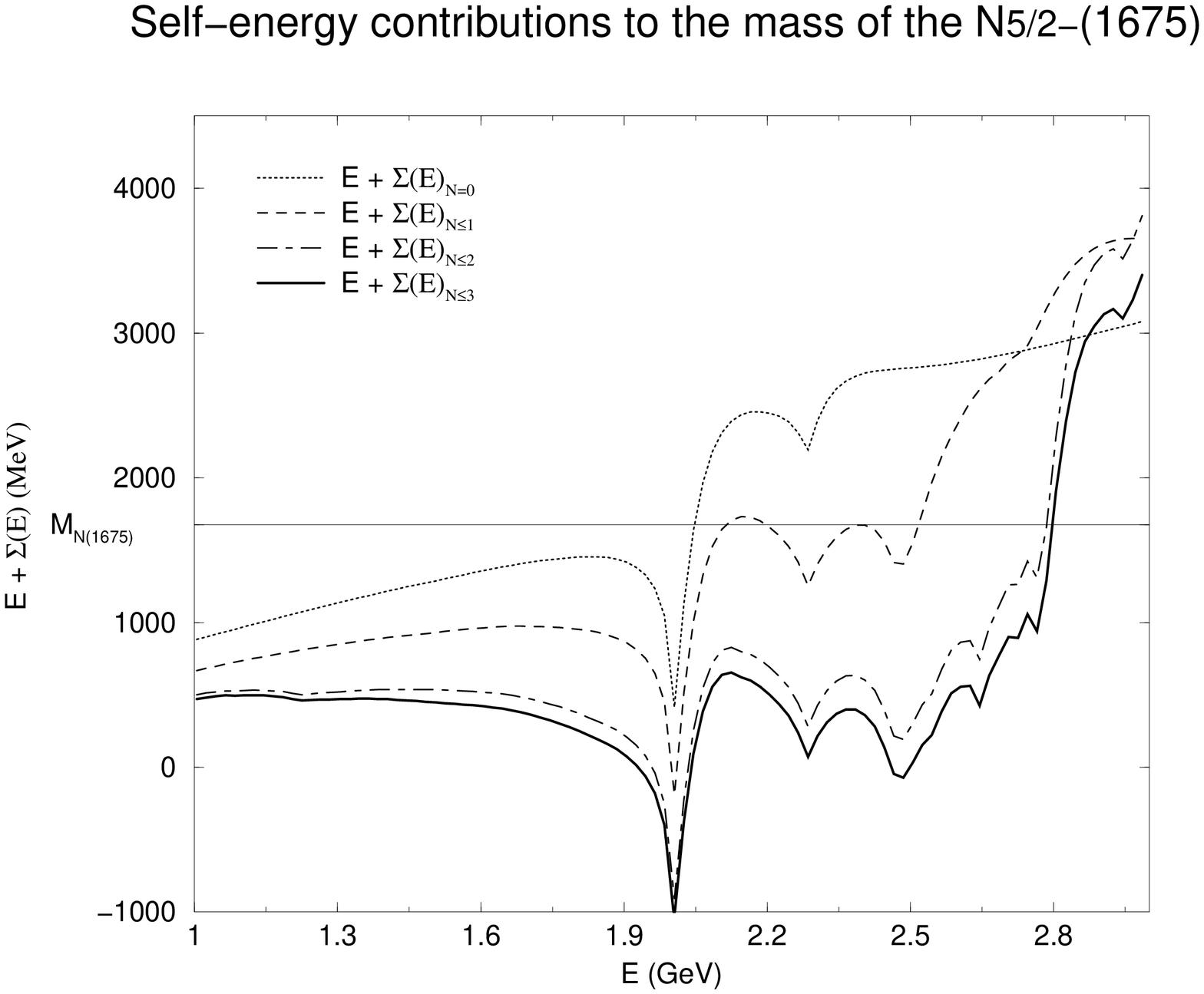,width=5.5in,height=6in,angle=270}
\caption[Self-energy contributions to the mass of
 $N\fhalf^-(1675)$]{\label{N5m1}Sum of the bare energy and self
 energies as a function of the bare energy for $N\fhalf^-(1675)$ with
 $\alpha_s=0.55$ and $\alpha=0.5$~GeV.}
\end{center}
\end{figure}

The results for the $L=1$ negative-parity states are presented in
Figures~\ref{N1m1_N1m2} through~\ref{N5m1} and
Table~\ref{halfat_4}(b). States with same quantum numbers are shown
together to facilitate comparisons.

Figure~\ref{N1m1_N1m2} shows the results for the spin-partner
$N\half^-$ states. It can immediately be seen that the bands of
intermediate states have a different impact on these states (and in
fact on all the $L=1$ negative-parity states) compared to the
situation with the Nucleon and $\Delta$ shown above. Here,
intermediate baryon states up to the $N=2$ band make sizeable
contributions but $N=3$ states only change the results
marginally. This indicates that the results have converged and that
all intermediate baryons up to and including the $N=2$ band states are
required for convergence.

If only ground-state baryons are included, the splitting between the
two $N\half^-$ states can be seen to be roughly 20 MeV, while it can
be observed to grow to roughly 100 MeV with the inclusion of the $N=1$
band baryons. Further addition of the $N=2$ and $N=3$ band states
brings convergence and closes this gap to about 5 MeV. This
illustrates the wide difference in results that can be obtained if the
set of intermediate states is not large enough to attain convergence.

Figure~\ref{N3m1_N3m2} shows similar results for the $N\thalf^-$
pair. Again the splitting between these states induced by these
self-energy effects varies from 25 MeV to 170 MeV depending on number
of intermediate states included. Note that in the case of the
$N\thalf^-(1520)$, although the $N=3$ band intermediate states were
not required for convergence, addition of these states resolved the
multiple solution problem. This validates the considerable extra
effort required to include such a large number of intermediate
states. Note again the differences in the threshold pattern between
the two states hinting at how differently these states couple to the
various intermediate states.

Figures~\ref{D1m1_D3m1} and ~\ref{N5m1} complete the set of
non-strange $L=1$ negative-parity baryon states with the
$\Delta\half^-(1620)$, $\Delta\thalf^-(1700)$, and $N\fhalf^-(1675)$
states. The two $\Delta$ states could actually be seen to be split by
a fair amount if only the $N=0$ and $N=1$ band intermediate states are
included, in a way mimicking spin-orbit splitting. However, with the
addition of the other two bands of states, the $\Delta(1620)$, which
was up to that point much lighter than its spin partner, becomes
almost degenerate with the $\Delta(1700)$ and even heavier by roughly
12 MeV. Of course these are states that are known to be affected
strongly by spin-orbit interactions between the quarks (which are not
included in this work), so the ordering is, at this time,
inconclusive. This stresses once again the impact of the choice of
intermediate states on the final results.
Since the graphs show robust results, we are confident that, for this
model, the final ordering is the correct one prior to the inclusion of
spin-orbit effects.
\subsection{$N=2$ Band States}
\begin{figure}
\begin{center}
\epsfig{file=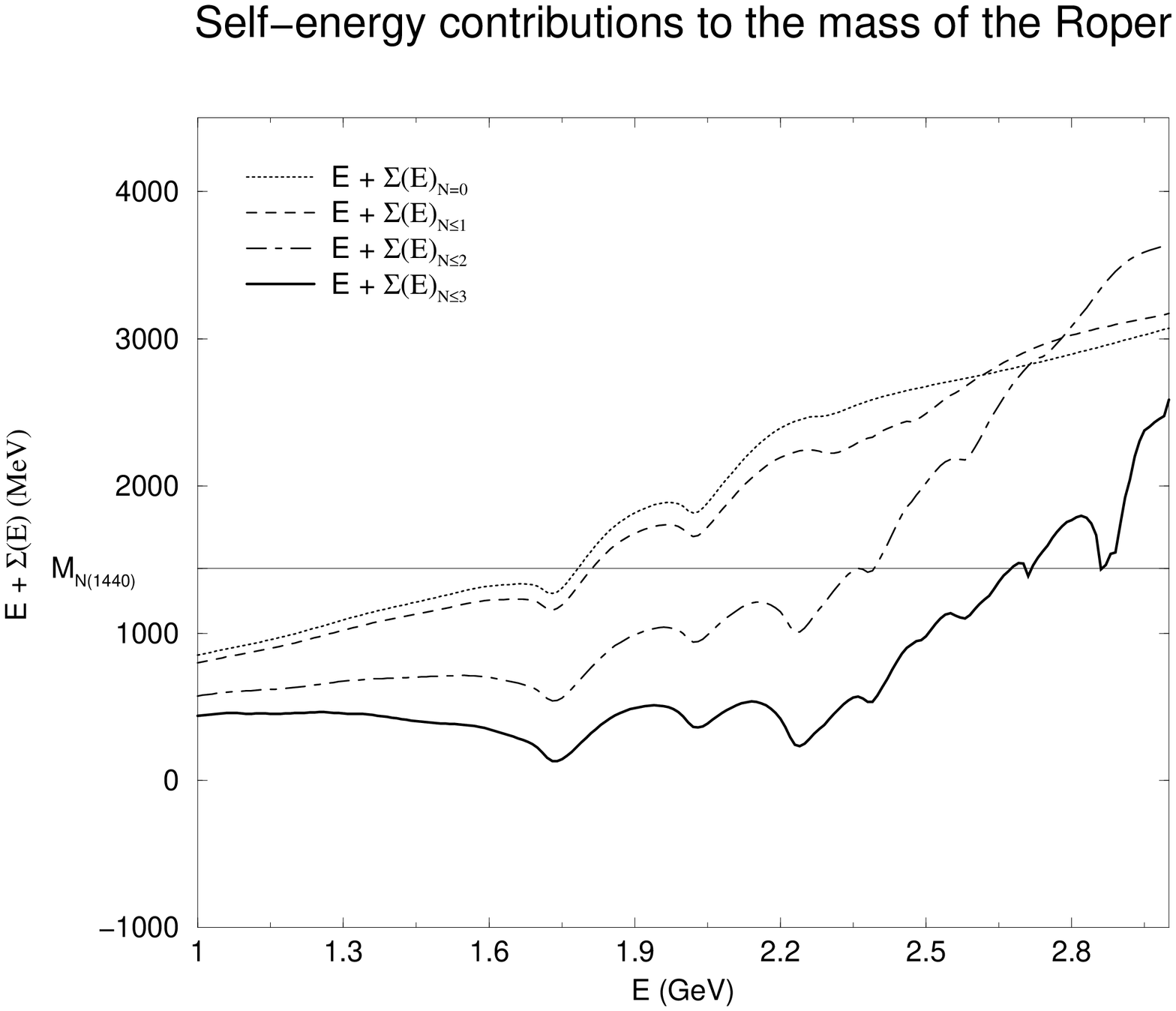,width=5.5in,height=6in,angle=270}
\caption[Self-energy contributions to the mass of 
$N\half^+(1440)$]{\label{N1p2}Sum of the bare energy and self energies
as a function of the bare energy for $N\half^+(1440)$ with
$\alpha_s=0.55$ and $\alpha=0.5$~GeV.}
\end{center}
\end{figure}
Finally, to illustrate a case where convergence is not yet achieved,
Figure~\ref{N1p2} is included to show the results for the Roper
resonance $N\half^+(1440)$, with these same set of intermediate
states. Not only do we not have a unique solution, but it is clear from
the difference between the bare mass for $N\leq 2$ and $N\leq 3$
baryon intermediate states included in the sum that the addition of
more intermediate states could significantly change the final
result. This indicates the need to extend the summation over
intermediate states to include at least $N=4$ band positive parity
baryon states. Note that the effect of the $N=1$ band intermediate
states is minimal, in contrast to the different situations shown
earlier. As expected, this indicates that as states from higher
harmonic oscillator bands are studied, inclusion of a large set of
intermediate states will be required, with a decreased impact of
the lower band states and an increased impact of the higher band
states.
\section{Hamiltonian vs. Self Energies}
As mentioned previously, the splittings between the states resulting
from the differences in self energies are expected to be comparable in
size to those that arise from residual interactions between quarks. As
a consequence, a self-consistent calculation requires that those
interactions be adjusted, and with them the wave functions, to account
for the additional splittings. A priori, we do not know how to modify
the Hamiltonian to get the desired result of agreement between the
bare energies and the modified $qqq$ spectrum, as each term affects
both the model masses and the baryon wave functions. The latter affect
the corresponding strong decay matrix elements and hence the size of
the self energies. To better understand this process, the splittings
of the states under consideration are examined using several different
values of the parameters listed in Table~\ref{Param}. Since the best
value of the $\tp0$ decay strength parameter $\gamma$ is affected by
changes in the wave functions, it was refitted each time to
reproduce the strength of the $\Delta\rightarrow N \pi$ decay
calculated with the model of Ref~\cite{CaRo:1993}. Additionally, the
harmonic oscillator parameter $\alpha$ is chosen on a coarse grid to
be such that the masses are roughly minimized, with the ground state
$\Delta$ being at or near its physical mass.

Four different cases are presented below; first, all quark-quark
residual interactions are turned off leaving only the confining
potential to act between quarks; second, the contact part of the
one-gluon exchange hyperfine interaction is included but at about half
the strength of the value used in Ref.~\cite{CaRo:1993}. Note that
although the value of $\alpha_s$ is only marginally lower (see
Table~\ref{Param}), lowering the value of $\sigma_0$ has the effect of
increasing the size of the quarks thereby reducing the strength of
short-range interactions. In the last two cases, the tensor part is
also included (proportionally to the contact interaction) and results
are presented for two different values of the oscillator parameter
$\alpha$. Each table shows the experimental masses of the states, the
model mass obtained from diagonalisation of the Hamiltonian, and the
bare masses, extracted from graphs similar to those presented above,
solving Eq.~\ref{selfe} for progressively larger sums of intermediate
states. The presence of multiple solutions is indicated by a range of
energies, or an additional value in parentheses when one answer was
favored.
\begin{table}
\caption[Bare masses with no residual quark
interactions.]{\label{zeroa}Bare masses (GeV) for $\alpha_s=0.0$ and
$\alpha=0.4$~GeV, no residual quark interactions.}
\vspace{0.25cm}
\begin{center}
\begin{tabular}{lccccc}
\hline\hline
\multicolumn{1}{c}{State}
& \multicolumn{1}{c}{Model}
& \multicolumn{1}{c}{$N=0$}
& \multicolumn{1}{c}{$N\leq 1$}
& \multicolumn{1}{c}{$N\leq 2$}
& \multicolumn{1}{c}{$N\leq 3$}\\
(Expt. Mass) & Mass & & & & \\
\hline
$[N\half^+](0.938)$      & 1.230 & 1.858 & 2.358 & 2.367 & 2.387-2.492 \\[6pt]
$[\Delta\thalf^+](1.232)$& 1.232 & 2.132 & 2.508 & 2.525 & 2.538 \\[6pt]
$[N\half^-](1.535)$      & 1.545 & 1.762 & 2.500 & 2.767 & 2.783 \\[6pt]
$[N\half^-](1.650)$      & 1.546 & 1.800 & 2.487 & 2.767 & 2.783  \\[6pt]
$[\Delta\half^-](1.620)$ & 1.546 & 1.812 & 2.467 & 2.767 & 2.787 \\[6pt]
$[N\thalf^-](1.520)$     & 1.546 & 1.758 & 2.400 & 2.787 & 2.800  \\[6pt] 
$[N\thalf^-](1.700)$     & 1.547 & 1.850 & 2.537 & 2.767 & 2.783  \\[6pt] 
$[\Delta\thalf^-](1.700)$& 1.546 & 2.037 & 2.550 & 2.878 & 2.800 \\[6pt]
$[N\fhalf^-](1.675)$     & 1.547 & 2.042 & 2.525 & 2.800 & 2.817  \\[6pt]
\hline\hline
\end{tabular}
\end{center}
\end{table}

When wave functions with no residual quark-quark interactions are used
to calculate the self energies, this results in the range of bare
masses shown in Table~\ref{zeroa}. As can be expected, model states
are mostly degenerate but this still results in a splitting between
the ground state $N$ and $\Delta$ of about 150 MeV due to the
difference in self energies from the $B'M$ loops. This splitting comes
from flavor and spin structure differences between those states, and
from differences in how each state couples to the various intermediate
states included in the sum. It was verified that if the masses of all
baryons and mesons are assumed to be degenerate, and only ground state
intermediate baryons are included in the sum, the splitting between
the $N$ and $\Delta$ ground states disappears, thereby verifying in
this model \.{Z}enczykowski~\cite{Ze:1986}'s statement about the minimum
number of intermediate baryon and meson states to be included to reach
this symmetry limit.
\begin{table}
\caption[Bare masses with hyperfine contact interactions
only.]{\label{halfa}Bare masses (GeV) for $\alpha_s=0.55$,
$\alpha=0.4$~GeV, with hyperfine contact interactions only.}
\vspace{0.25cm}
\begin{center}
\begin{tabular}{lccccc}
\hline\hline
\multicolumn{1}{c}{State}
& \multicolumn{1}{c}{Model}
& \multicolumn{1}{c}{$N=0$}
& \multicolumn{1}{c}{$N\leq 1$}
& \multicolumn{1}{c}{$N\leq 2$}
& \multicolumn{1}{c}{$N\leq 3$}\\
(Expt. Mass) & Mass & & & & \\
\hline
$[N\half^+](0.938)$      & 1.081 & 1.812 & 2.342        & 2.358 & 2.375  \\[6pt]
$[\Delta\thalf^+](1.232)$& 1.232 & 2.112 & 2.500        & 2.508 & 2.533  \\[6pt]
$[N\half^-](1.535)$      & 1.505 & 1.742 & 2.108        & 2.775 & 2.787  \\[6pt]
$[N\half^-](1.650)$      & 1.588 & 1.787 & 2.475        & 2.750 & 2.758  \\[6pt]
$[\Delta\half^-](1.620)$ & 1.568 & 1.787 & 2.200(2.425) & 2.775 & 2.787  \\[6pt]
$[N\thalf^-](1.520)$     & 1.505 & 1.750 & 2.0625       & 2.612 & 2.700-2.787  \\[6pt] 
$[N\thalf^-](1.700)$     & 1.588 & 1.787 & 2.358(2.512) & 2.758 & 2.775  \\[6pt] 
$[\Delta\thalf^-](1.700)$& 1.568 & 2.033 & 2.358(2.525) & 2.787 & 2.792  \\[6pt]
$[N\fhalf^-](1.675)$     & 1.588 & 2.037 & 2.100(2.500) & 2.733(2.775) & 2.787 \\[6pt]
\hline\hline
\end{tabular}
\end{center}
\end{table}

Table~\ref{halfa} shows the changes created by the inclusion of the
contact part of the hyperfine interaction (with no tensor or
spin-orbit interactions). As mentioned above, the contact interaction
is roughly half the strength of that used in Ref.~\cite{CaRo:1993} and
the following papers. In the model, the degeneracy is now lifted
between the spin partners $N$-$
\Delta$, the $N\half^-$ states, and $N\thalf^-$ states. Interestingly,
the spin-orbit partners $\Delta\half^-$ and $\Delta\thalf^-$ states
remain degenerate. Note that spin-orbit interactions are not included
in their wave functions. It is important to note that after the
addition of all intermediate $B'M$ states, the $N$-$\Delta$ splitting
remains roughly 150 MeV, demonstrating a stable result for these
states. Note also that the ordering of the $N\half^-$ states is seen
to change as more intermediate states are added, emphasizing the point
that not all relevant effects have been included by previous
calculations which include only $N=0$ intermediate states.
\begin{table}
\caption[Bare masses with hyperfine contact and tensor
interactions.]{\label{halfat_4}Bare masses (GeV) for $\alpha_s=0.55$
 with hyperfine contact and tensor interactions.}
\vspace{0.25cm}
\begin{center}
{\bf a)} $\alpha=0.4$~GeV.\\
\vspace{0.5cm}
\begin{tabular}{lccccc}
\hline\hline
\multicolumn{1}{c}{State}
& \multicolumn{1}{c}{Model}
& \multicolumn{1}{c}{$N=0$}
& \multicolumn{1}{c}{$N\leq 1$}
& \multicolumn{1}{c}{$N\leq 2$}
& \multicolumn{1}{c}{$N\leq 3$}\\
(Expt. Mass) & Mass & & & &\\
\hline
$[N\half^+](0.938)$       & 1.082 & 1.812 & (2.175)2.342 & 2.358 & 2.375 \\[6pt]
$[\Delta\thalf^+](1.232)$ & 1.232 & 2.112 & (2.358)2.500 & 2.508 & 2.525 \\[6pt]
$[N\half^-](1.535)$       & 1.500 & 1.745 & 2.105        & 2.712 & 2.735 \\[6pt]
$[N\half^-](1.650)$       & 1.572 & 1.783 & 2.412        & 2.737 & 2.758 \\[6pt]
$[\Delta\half^-](1.620)$  & 1.570 & 1.783 & 2.200(2.412) & 2.725(2.775) & 2.787 \\[6pt]
$[N\thalf^-](1.520)$      & 1.506 & 1.750 & 2.062        & 2.600 & 2.650-2.785 \\[6pt] 
$[N\thalf^-](1.700)$      & 1.606 & 1.787 & 2.350(2.512) & 2.775 & 2.812 \\[6pt] 
$[\Delta\thalf^-](1.700)$ & 1.569 & 2.375 & 2.350(2.530) & 2.787 & 2.795 \\[6pt]
$[N\fhalf^-](1.675)$      & 1.584 & 2.037 & 2.100(2.492) & 2.787 & 2.800 \\[6pt]
\hline\hline
\end{tabular}
\end{center}

\begin{center}
{\bf b)} $\alpha=0.5$~GeV.\\
\vspace{0.5cm}
\begin{tabular}{lccccc}
\hline\hline
\multicolumn{1}{c}{State}
& \multicolumn{1}{c}{Model}
& \multicolumn{1}{c}{$N=0$}
& \multicolumn{1}{c}{$N\leq 1$}
& \multicolumn{1}{c}{$N\leq 2$}
& \multicolumn{1}{c}{$N\leq 3$}\\
(Expt. Mass) & Mass & & & & \\
\hline
$[N\half^+](0.938)$        & 1.082 & 1.850        & 2.367        & 2.375 & 2.392(2.500)\\[6pt]
$[\Delta\thalf^+](1.232)$  & 1.232 & 2.137        & 2.508        & 2.517 & 2.537 \\[6pt]
$[N\half^-](1.535)$        & 1.500 & 1.762        & 2.375        & 2.737 & 2.758 \\[6pt]
$[N\half^-](1.650)$        & 1.572 & 1.783        & 2.475        & 2.742 & 2.762 \\[6pt]
$[\Delta\half^-](1.620)$   & 1.570 & 1.800(2.025) & 2.467        & 2.800 & 2.812\\[6pt]
$[N\thalf^-](1.520)$       & 1.506 & 1.762        & 2.082(2.367) & 2.637(2.793) & 2.800\\[6pt] 
$[N\thalf^-](1.700)$       & 1.606 & 1.787        & 2.537        & 2.812 & 2.825 \\[6pt] 
$[\Delta\thalf^-](1.700)$  & 1.569 & 2.050        & 2.558        & 2.787 & 2.800 \\[6pt]
$[N\fhalf^-](1.675)$       & 1.584 & 2.050        & (2.100)2.512 & 2.787 & 2.800 \\[6pt]
\hline\hline
\end{tabular}
\end{center}
\end{table}

Tables~\ref{halfat_4} show the results (previously illustrated in
Figures~\ref{Nuc} through~\ref{N5m1}) when both the contact and tensor
parts of the hyperfine interaction are included in the
Hamiltonian. Here the tensor interaction has the strength required
from a consistent nonrelativistic limit of one-gluon exchange. Each
table reflects a different value of the harmonic oscillator parameter
$\alpha$ (0.4 and 0.5 GeV). Again the $N$-$\Delta$ splitting is
unaffected by the changes, as the wave functions for these two states
are essentially unchanged by this change in the basis states. The
splittings in the bare masses of other states are not strongly affected
by the change in $\alpha$. The model masses are minimized with a value
of $\alpha=$0.5 GeV, so the results using this basis are preferred.
\section{The Spectrum}
Finally this section is brought to a close with the presentation of
three figures showing the progression of the relationship between the
model masses and the bare masses required to reproduce the masses of
the states extracted from data analyses. In each figure the two
different mass scales have been adjusted so that the bare and model
masses of the ground state $\Delta$ coincide. The bare masses shown
include all intermediates $B'M$ states for each set of parameters.
\begin{figure}
\begin{center}
\epsfig{file=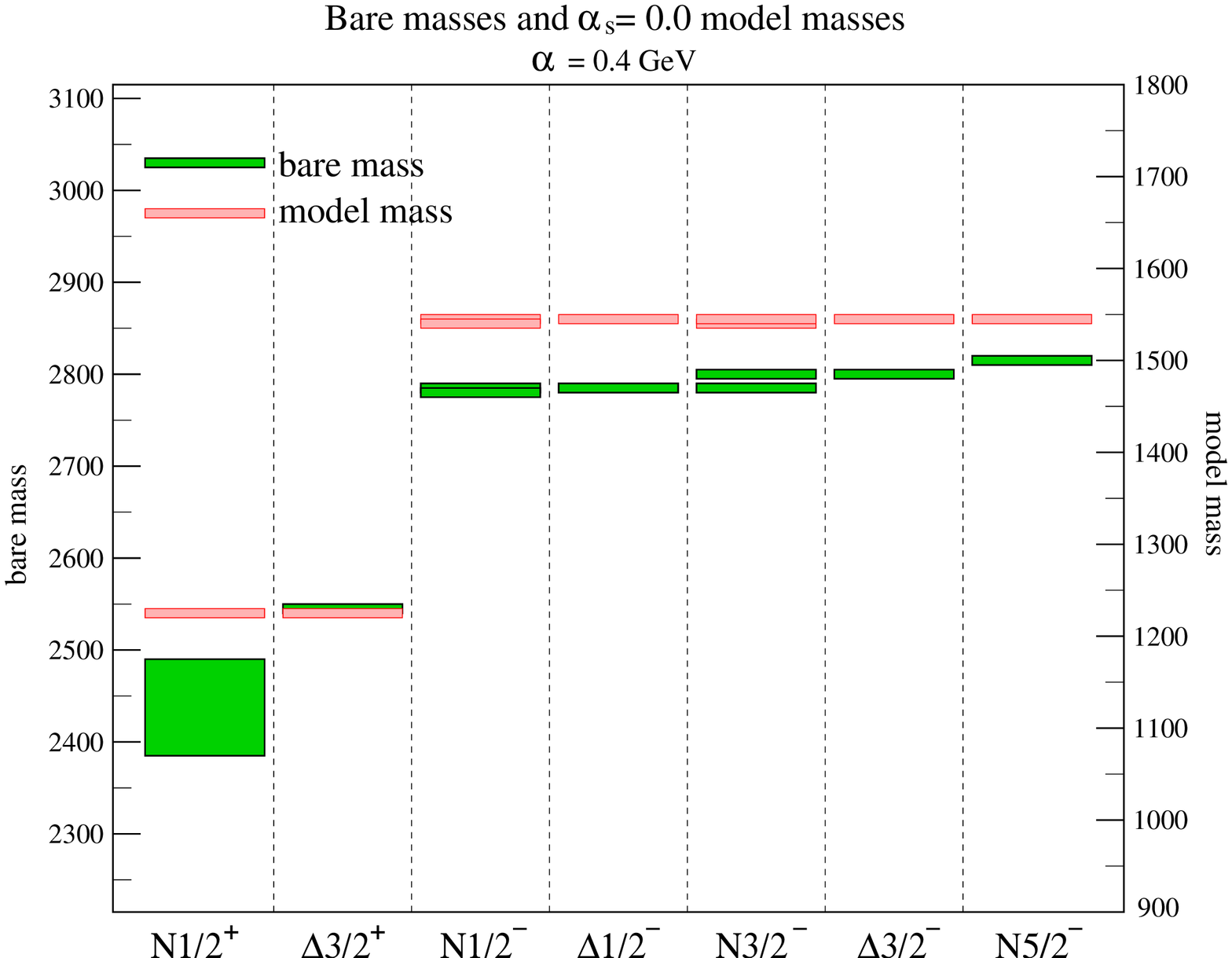,height=5in,width=6in,angle=0}
\caption[Bare masses versus Model masses with no residual
interactions]{\label{zerospectrum}Comparison of the spectra of the
bare masses required to fit the physical masses of the states shown
and model masses obtained from a Hamiltonian with no residual
quark-quark interactions. Here $\alpha_s=0.0$ and $\alpha=0.4$~GeV.}
\end{center}
\end{figure}

When all residual interactions between quarks have been removed, the
spectrum of the states studied appears as shown in
Figure~\ref{zerospectrum}. As stated above and seen here, model masses
are degenerate by oscillator band. The inclusion of self energy loops
requires different bare energies to fit the physical masses of the $N$
and $\Delta$ ground states, and reduces the splitting in the bare
energies between oscillator bands. The Hamiltonian therefore produces
model masses in bands that are too far apart at this point.
\begin{figure}
\begin{center}
\epsfig{file=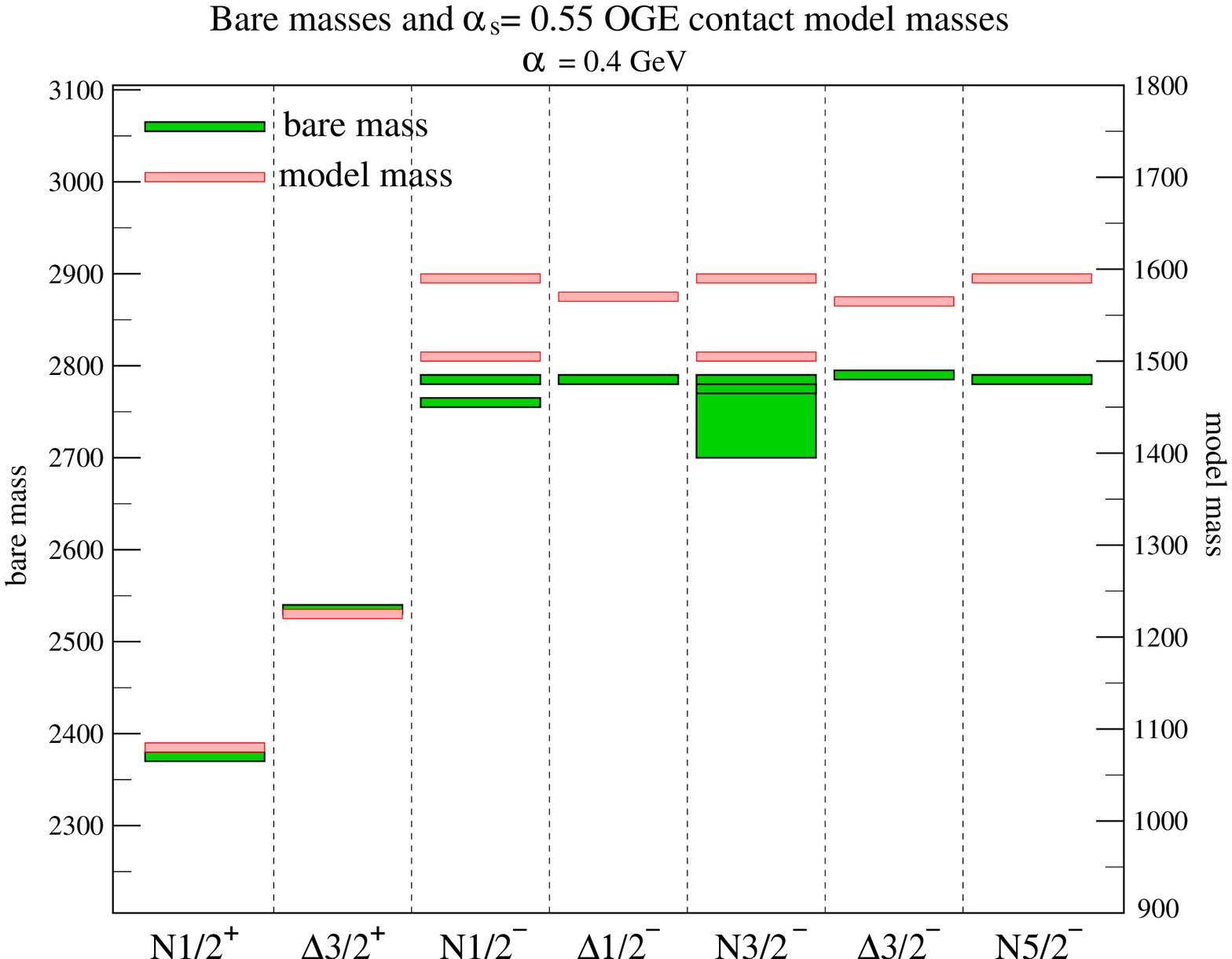,height=5in,width=6in,angle=0}
\caption[Bare masses versus Model masses with hyperfine contact
interactions only]{\label{halfspectrum}Comparison of the spectra of
the bare masses required to fit the physical masses of the states
shown and model masses obtained from a Hamiltonian with hyperfine
contact interactions only. Here $\alpha_s=0.55$ and $\alpha=0.4$~GeV.}
\end{center}
\end{figure}

As seen in Figure~\ref{halfspectrum}, the addition of reduced-strength
contact interactions between quarks induces configuration mixing in
the wave functions and lifts the degeneracy between spin
partners. States split by other type of interactions remain mostly
unchanged by this addition.  The effect of the self energies on the
bare masses extracted from the physical masses also changes, although
the $N$-$\Delta$ bare mass splitting is stable at roughly 150 MeV. The
order of the two $N\half^-$ states, on the other hand, is reversed
with the bare mass of the predominantly spin-1/2 state heavier than
that of the predominantly spin-3/2 state, and the splitting is
larger. The splitting between the bare masses of $N\thalf^-$ states
appears to be reduced by the inclusion of loops but the presence of
multiple solutions blurs the picture somewhat.
\begin{figure}
\begin{center}
\epsfig{file=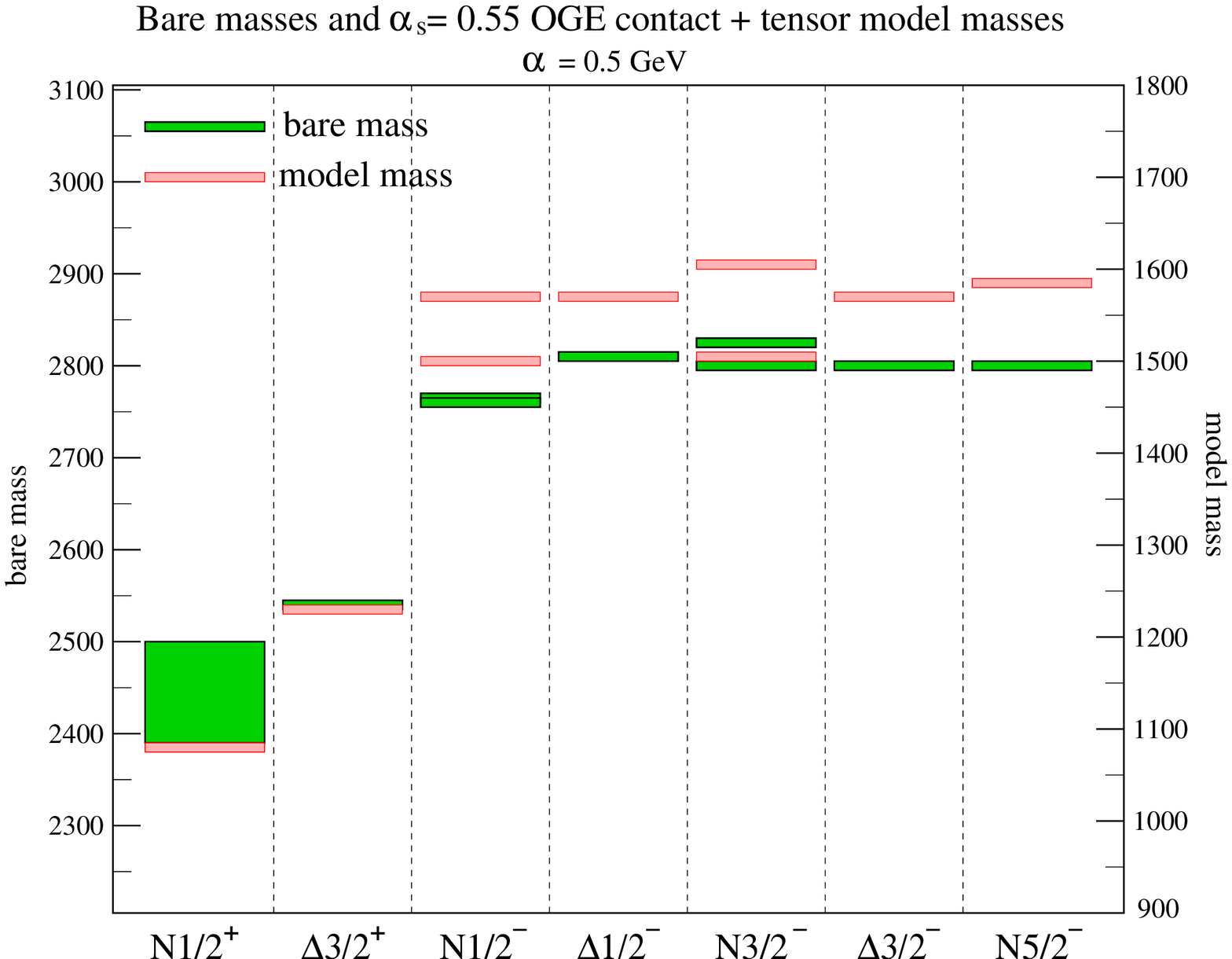,height=5in,width=6in,angle=0}
\caption[Bare masses versus Model masses with hyperfine contact and
tensor interactions]{\label{halftspectrum}Comparison of the spectra of
the bare masses required to fit the physical masses of the states
shown and model masses obtained from a Hamiltonian with hyperfine
contact and tensor interactions. Here $\alpha_s=0.55$ and
$\alpha=0.5$~GeV.}
\end{center}
\end{figure}

The addition of the tensor part of the hyperfine interaction changes
the spectra once again, as shown in Figure~\ref{halftspectrum}. The
modifications to the Hamiltonian close slightly the splitting between
the model masses of the $N\half^-$ states, and open it for the
$N\thalf^-$ states. On the other hand the required bare energies of the
$N\half^-$ states are almost degenerate, and those of the $N\thalf^-$
states are separated by about 25 MeV. It is expected that the mixing
between states with same quantum numbers due to self energy loops will
widen the gaps in both cases, so this picture is not yet
complete. The splitting between oscillator bands is still larger for
the model spectrum than for the bare mass spectrum but the overall
agreement between the spectra remains fairly good.

In closing, it is important to mention the extensive computational
work required to obtain the results presented in this Chapter and
which comprised the most time consuming part of this work. The code
used to produce the analytical form of the momentum dependence of the
strong decay vertices was entirely done using the symbolic manipulator
{\it Maple}. The code is based on the general method of Roberts and
Silvestre-Brac~\cite{RoSB:1992} and was thoroughly tested by
reproducing the large number of decay amplitudes found in several
published papers by Capstick and
Roberts~\cite{CaRo:1993,CaRo:1994,CaRo:1998}. The analytical
expressions produced by the {\it Maple} code were subsequently
translated into the programming language {\it C}, and then included in
the code which numerically calculates the principal part of the loop
integration using algorithms such as Gauss-Laguerre and Gauss-Legendre
quadratures. These extensive calculations would not have been possible
without access to the FSU Physics Department Computing Cluster. The
computational time currently required to obtain the results listed in
just {\bf one} of the tables presented above is of the order of 5 days
of full-time computing for an average of 10 nodes in the cluster. This
does not, however, reflect the many months of intensive computing work
done prior to this `step' when several of the analytical components of
the decay amplitudes were computed and stored. The interested reader
is referred to Appendix~\ref{ap:COMP} for more details about the
computational methods used during this project.

\chapter{Conclusion}
\label{Concl}
Within the context of the relativized quark model, a comprehensive
study has been carried out of the effects of baryon-meson intermediate
states, in the form of self-energy loop corrections to the mass of
several baryons. The baryon states whose self energies have been
studied include the nucleon and Delta ground states and the first band
of negative-parity excited baryons. The $\tp0$ decay model is used to
obtain the analytical form of the momentum dependence of the
baryon-baryon-meson vertices needed in the loop calculations. The model
is modified to take into account the size of the constituent-quark
pair-creation vertex. Wave functions generated from a Hamiltonian
including reduced-strength one-gluon-exchange interactions between
quarks are used to calculate the self energies. Since masses play a
crucial role in the size of the self energies, physical masses are
used where known, and model masses~\cite{CaRo:1993} used otherwise,
for the intermediate baryons and mesons. The bare energy of the
initial baryon is determined self-consistently by calculating the self
energies for a range of bare energies, then finding the solution to
Eq.~\ref{selfe} when $M_B=M_{\rm physical}$ for each initial baryon
studied.

As demonstrated by \.{Z}enczykowski~\cite{Ze:1986}, a minimum set of
baryon-meson intermediate states is required to recover the
SU(3)$_{f}\times$SU(2)$_{\rm spin}$ symmetry limit, while Brack and
Bhaduri~\cite{BrBh:1987} showed that intermediate baryon states up to
at least the second band ($N=3$) of negative-parity excited states
must be included in order for the sum over intermediate baryon-meson
states to converge. It is clear that truncation at only ground state
baryons or only pseudo-scalar mesons leads to physically meaningless
results. The present work therefore uses for the first time a complete
set of spin-flavor symmetry related baryon-meson intermediate states,
while at the same time including excited baryon states up to the $N=3$
band, to insure the convergence of the sum of intermediate states. It
is shown that drastically different answers are obtained for some
states if this sum is indeed truncated at smaller sets of intermediate
states, and that convergence is reached for all the external baryon
states considered in this work. It is also shown that a larger number
of intermediate states is required to reach convergence as the initial
state becomes more highly excited. For example, the set of
intermediate states included in this work is found to be insufficient
for initial states in the first ($N=2$) positive-parity excited state
band, such as the Roper resonance. This and other states in the same
harmonic oscillator band will require the inclusion of baryon
intermediate states up to at least the second ($N=4$) positive-parity
band.

The existence of decay thresholds and their effects on the self
energies are shown. Since physical masses are used where available,
actual thresholds can be identified on the graphical version of some
of the results. Their presence creates an oscillatory pattern in the
curves for the sum of the bare energy and the self energies, that may
lead to multiple solutions to Eq.~\ref{selfe}. Such situations become
less frequent as the sum over intermediate states is expanded and
convergence is reached.

In this model, it is found that roughly half of the splitting between
the nucleon and Delta ground states arises from self energy loop
effects, the other half coming from residual quark-quark
interactions. Changes in these interactions have very little impact on
this result since they do not affect the wave functions of these
states very strongly. The effects of the same set of intermediate
states on the spectrum of $L=1$ negative-parity excited states is also
examined, and it is found that the resulting splittings are sensitive
to configuration mixing in the baryon wavefunctions caused by residual
interactions between the quarks. Additionally, some of these states
are expected to mix further due to off-diagonal terms in their self
energies. Fairly good agreement is found between the spectrum of bare
masses produced by the inclusion of a large set of baryon-meson
intermediate states and the spectrum of model masses obtained from a
Hamiltonian with hyperfine and contact interactions between
quarks. The overall shift between the two spectra of negative-parity
states could be attributed to a problem with the string tension, which
has been shown~\cite{GeIs:1990} to be renormalized by the presence of
self-energy loops. Spin-orbit interactions are also expected to play
an important role in changing both model masses and bare masses for
the negative-parity excited states therefore the results of this work
can be considered a significant step toward an understanding of these
states but work remains to be done.
\begin{figure}
\begin{center}
\epsfig{file=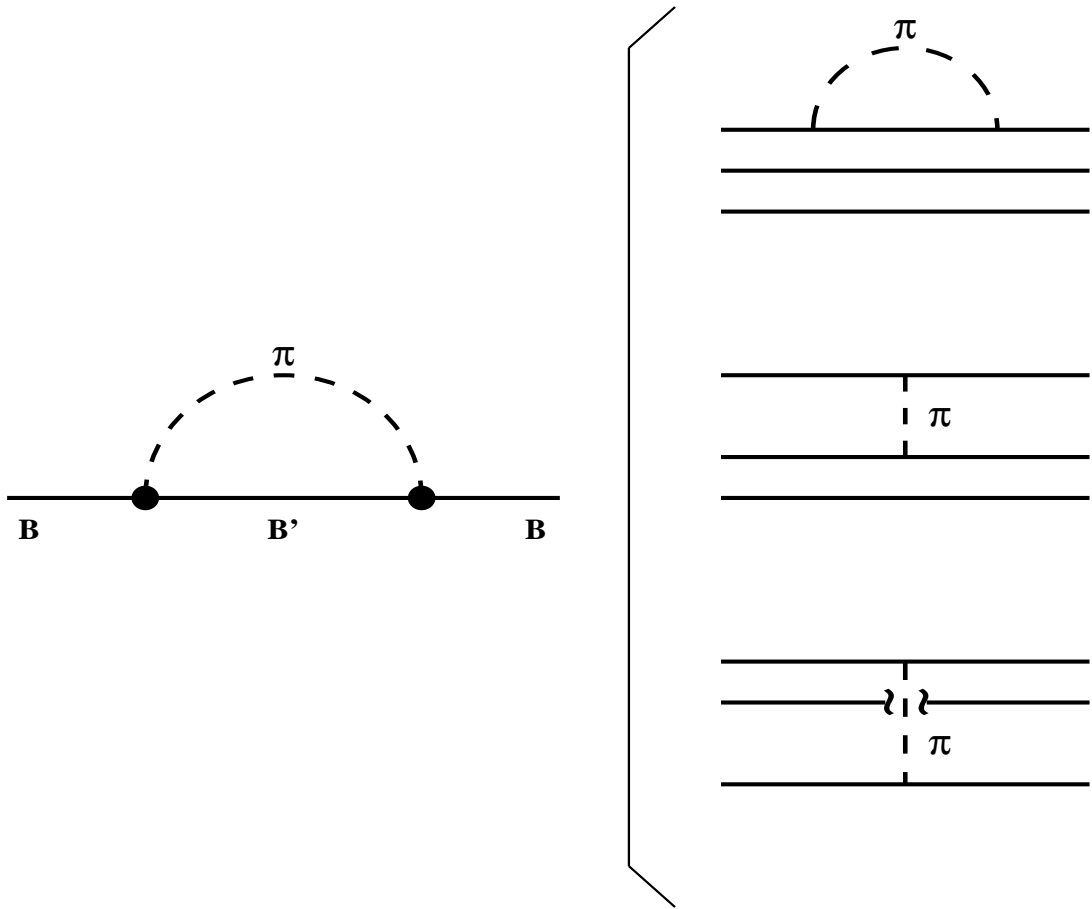,height=5in,width=6in,angle=0}
\caption[Self-energy Loops and
One-Boson-Exchange]{\label{model}Self-energy model includes both loop
effects and one-boson-exchange effects in a self-consistent
calculation}
\end{center}
\end{figure}

It is important to note that the model used in this work not only
self-consistently incorporates the effects on the properties of the
quarks of self-energy loops, but also the effects of
one-boson-exchange mechanism on baryon properties, since the created
anti-quark can merge with any quark from the initial and final
baryon to produce the intermediate meson. This point is illustrated in
Figure~\ref{model}. Calculations which treat only short distance
interactions between quarks from the high momentum transfer limit of
the last two diagrams on the right hand side of Figure~\ref{model} can
be expected to neglect important effects from the self-consistent
evaluation of quark self energies and exchange effects due to all
mesons. They also do not properly take into account the two-hadron
nature of the intermediate state at low momentum scales.

As mentioned before, and as it should also be apparent from the
results of this work, conclusions made in prior works about spin-orbit
forces in negative-parity excited baryon states were premature. Any
calculation not including a complete set of spin-flavor symmetry
related baryon-meson intermediate states, {\bf and} excited baryon
states up to at least the $N=3$ band, cannot claim to have complete
results.
\section{Outlook}
After extensive work on this project there still remain many
unanswered questions and unaddressed concerns. One important asset of
the computational tools assembled for this project is their ability to
be extended to larger sets of initial and intermediate hadrons given
sufficient time and computing resources. Below, a few of the projects
lined up behind this one are touched upon.

Since the framework is now in place, the next step is extension to the
strange sector. The ground state $\Lambda$ and $\Sigma$ baryons as
well as some of the experimentally better known $L=1$ negative-parity
band $\Lambda$ and $\Sigma$ states are already under
study. Experimental data for several of these states has recently
improved, but agreement between the analyses of this experimental data
and model predictions for resonance parameters is far from perfect. It
will be interesting to find out how the baryon-meson loop induced self
energies affect the splittings within this set of states.

As pointed out earlier, the inclusion of at least $N=4$ band baryon
intermediate states will be required to bring about convergence of the
intermediate state sum for $N=2$ band initial states. This should
shed some light on the nature of some controversial and hard to model
positive-parity states such as the Roper resonance. The computational
tools used in this work are easily extended to such a calculation.

Mixing due to self energies must contribute to the splitting between
baryons with same quantum numbers. Therefore studies of the mixing
between the Nucleon and the Roper, the $N\half^-(1535)$ and
$N\half^-(1650)$ states, and the $N\thalf^-(1520)$ and $N\thalf^-(1700)$
states will be required to further understand the impact of self energy
loops on the splittings of these pairs of states. 

The addition of spin-orbit interactions needs to be further
explored. A preliminary study shows that the addition of spin-orbit
interactions in the interactions leading to the wave functions used for
the decay vertices can change the splitting between some states and
even reverse the ordering of some of the states. More work is clearly
needed in this area before conclusions can be drawn, as both the
mixings due to self energies and spin-obit interactions strongly
affect the masses of these states.

Even with these cautions, the present calculation goes far beyond
anything previously available for the negative-parity non-strange
baryons and shows, for the first time, results with a set of
intermediate states large enough to achieve convergence. It also
demonstrates for the first time the sensitivity of the self energies
to the mixings caused by various components of the quark residual
interactions.
\appendix           
\chapter{The Wave Functions}
\label{ap:WF}
 
There are four components in each hadron wave function: color ($C$),
flavor ($\phi$), spin ($\chi$), and spatial ($\psi$) wave functions.
\section{Color}
The meson color wave function is found by the rules of $SU(3)_c$ for
direct products of quarks ($q$) and antiquarks ($\bar{q}$) carrying
color charges or ${\bf 3} \otimes {\bf \bar{3}} = {\bf 1} \oplus {\bf
8}$ and must be singlet to be an observable therefore it is
\begin{equation}
C^M = \sum_{i=1}^3 \  \frac{1}{\sqrt{3}} \ q_i \ \bar{q}_i.
\end{equation} 

Similarly, the baryon color wave function is found by the same rules but
for three quarks or ${\bf 3} \otimes {\bf 3} \otimes {\bf 3} = {\bf 1}
\oplus {\bf 8} \oplus {\bf 8'} \oplus {\bf 10}$ to be the totally
antisymmetric singlet combination
\begin{equation}
C_A^B = \sum_{i,j,k=1}^3 \  \frac{1}{\sqrt{6}}\ \epsilon_{ijk} \ q_1^i \
q_2^j \  q_3^k.
\end{equation}
\section{Flavor}
The flavor wave functions $\phi$ for the baryons and mesons included in this
calculation are shown in Table~\ref{flav}. They are obtained
from the irreducible representations of $SU(3)_F$ giving a flavor
nonet each for the pseudoscalar mesons ($\pi,K,\bar {K},\eta,\eta'$) and the
vector mesons ($\rho,K^*,\bar {K}^*,\omega,\phi$)\footnote{Note that
since the $\phi$ mesons couple weakly to non-strange baryon states (such decays
are OZI suppressed), they are not included.}. 

For baryons, we follow
the convention used in ref.~\cite{CaRo:1993} and adopt a generalized
$uds$ basis that only symmetrizes the product $\phi \chi \psi$ in {\it
identical} quarks. This removes the need for symmetrization between
the $u$ and $d$ quarks in the spatial wave function making it manageable
to work with states with up to $7\hbar \omega$ in the harmonic oscillator
spectrum. Note that the baryon flavor wave functions are all either
symmetric or antisymmetric under the interchange of quarks one and two.  
\begin{table}
\caption[Flavor wave functions.]{\label{flav}The baryon and meson
flavor wave functions as a function of their isospin projection ($I_z$).}
\vspace{0.5cm}
\begin{center}
\begin{tabular}{cccccccc}
\hline\hline
\multicolumn{1}{l}{State}
& \multicolumn{1}{c}{$+3/2$}
& \multicolumn{1}{c}{$+1$}
& \multicolumn{1}{c}{$1/2$}
& \multicolumn{1}{c}{$0$}
& \multicolumn{1}{c}{$-1/2$}
& \multicolumn{1}{r}{$-1$}
& \multicolumn{1}{c}{$-3/2$}\\
\hline
$N$ & & & $uud$ & & $ddu$ & &\\[4pt]
$\Delta$ &  $uuu$ & & $uud$ & & $ddu$ & & $ddd$\\[4pt]
$\Lambda$ & & & & $\frac{1}{\sqrt{2}}(ud-du)s$ & &\\[4pt]
$\Sigma$  & & $uus$ & & $\frac{1}{\sqrt{2}}(ud+du)s$ & & $dds$\\[4pt]
\hline \hline
$\pi$ & & $-u\bar{d}$ & & $\frac{1}{\sqrt{2}}(u\bar{u}-d\bar{d})$  & & $+d\bar{u}$ \\[6pt]
$\rho$ & &$-u\bar{d}$ & &  $\frac{1}{\sqrt{2}}(u\bar{u}-d\bar{d})$ & & $+d\bar{u}$ \\[6pt]
$K$ & & & $-u\bar{s}$ & & $+s\bar{u}$ & \\[6pt]
$K^0$ & & & $-d\bar{s}$ & & $-s\bar{d}$ & \\[6pt]
$\eta$ & & & & $\frac{1}{\sqrt{2}}\left[
\frac{1}{\sqrt{2}}(u\bar{u}+d\bar{d})-s\bar{s} \right]$ & & \\[6pt]
$\eta'$ & & & & $\frac{1}{\sqrt{2}}\left[
\frac{1}{\sqrt{2}}(u\bar{u}+d\bar{d})+s\bar{s} \right]$ & & \\[6pt]
$\omega$ & & & & $\frac{1}{\sqrt{2}}(u\bar{u}+d\bar{d})$ & &  \\[6pt]
\hline \hline
\end{tabular}
\end{center}
\end{table}
\section{Spin}
The total spin of two spin-$\half$ particles can be either zero or one
giving us the pseudoscalar and vector mesons respectively.

The total spin of the three spin-$\half$ particles can be either
$\half$ or $\thalf$ so that as a complete set of spin wave functions $\chi$ we can choose
\begin{eqnarray}
\chi^S_{{3\over 2}{3\over 2}} &=&
\vert \uparrow \uparrow \uparrow \ \rangle\ \ ,\ \ {\rm{etc.}} \\
\chi^{M_{\rho}}_{{1\over 2}{1\over 2}}&=&
{1\over \sqrt{2}}\left(\ \vert\uparrow\downarrow\uparrow\ \rangle -
\vert\downarrow\uparrow\uparrow\ \rangle \ \right)\ \ ,\ \ {\rm{etc.}} 
\\
\chi^{M_{\lambda}}_{{1\over 2}{1\over 2}} &=&
-{1\over \sqrt{6}}\left(\ \vert\uparrow\downarrow\uparrow\ \rangle +
\vert\downarrow\uparrow\uparrow\ \rangle
-2\vert\uparrow\uparrow\downarrow\ \rangle \ \right)\ \ ,\ \ {\rm{etc.}}
\end{eqnarray}
(We show only the top state of a JM multiplet; other 
wave functions follow the Condon-Shortley convention). Note that the baryon 
spin wave functions are also either symmetric or antisymmetric under 
interchange of the first two quarks.
\section{Space}
Finally, for the spatial wave functions $\Psi$ we take functions with
definite total ${\bf L}= {\bf l}_{\rho}+ {\bf l}_{\lambda}$ made from a
Clebsch-Gordan sum of harmonic oscillator wave functions in the two
relative coordinates 
\begin{eqnarray}
\lpmb{\rho} &\equiv& {1\over \sqrt{2}} ({\bf r}_1-{\bf r}_2)
\end{eqnarray}
and 
\begin{eqnarray}
\lpmb{\lambda} &\equiv& {1\over \sqrt{6}} ({\bf r}_1+{\bf r}_2-2{\bf r} _3)
\end{eqnarray}

\begin{figure}
\setlength{\unitlength}{0.5cm}
\hspace{2cm}
\begin{picture}(15,15)
\thicklines
\put(5,5){\circle{2}}
\put(15,7.5){\circle{2}}
\put(10,15){\circle{2}}
\put(14,7.5){\vector(-4,-1){8.1}}
\put(10,14){\vector(0,-1){7.5}}
\put(10,5.5){$\frac{\bf{r}_1 +\bf{r}_2}{2}$}
\put(10.5,10){$\frac{\sqrt{6}}{2}\lpmb{\lambda}$}
\put(7,6.5){$\sqrt{2}\lpmb{\rho}$}
\put(4.75,4.75){$1$}
\put(14.75,7.25){$2$}
\put(9.75,14.75){$3$}
\end{picture}
\vspace{-1cm}
\caption[Relative coordinates.]{\label{figB1}Relative coordinates
$\lpmb{\rho}$ and $\lpmb{\lambda}$.}
\end{figure}
\noindent of the three body problem (see Figure~\ref{figB1}). These
are
\newpage
\begin{eqnarray}
\Psi_{_{LMn_{\rho}l_{\rho}n_{\lambda}l_{\lambda}}}&=&\alpha^3\sum_m 
C(l_{\rho}\,l_{\lambda}\,m\,M-m\,;\,L\,M)
{\cal {N}}_{n_{\rho}l_{\rho}}(\alpha \rho)^{l_{\rho}}
e\,^{-{1\over 2}\alpha^2\rho^2}
L_{n_{\rho}}^{l_{\rho}+{1\over 2}}(\alpha \rho)
Y_{l_{\rho}m}(\Omega_{\rho}) \nonumber\\ 
&&\times  {\cal {N}}_{n_{\lambda}l_{\lambda}}(\alpha \lambda)^{l_{\lambda}}
e\,^{-{1\over 2}\alpha^2\lambda^2}
L_{n_{\lambda}}^{l_{\lambda}+{1\over 2}}(\alpha \lambda)
Y_{l_{\lambda}M-m}(\Omega_{\lambda}) ,
\end{eqnarray}
where the $L_n^{l+{1\over 2}}(x)$ are the associated Laguerre 
polynomials
\begin{equation}\label{Laguerre}
L_n^{l+{1\over 2}}(x)=\sum_{m=0}^n(-1)^m
{n+l+{1\over 2}\choose n-m}{x^{2m} \over m!}
\end{equation}
(half-integral factorials are defined by the $\Gamma$ function), and the 
normalization coefficient ${\cal {N}}_{nl}$ is defined by 
\begin{equation}\label{Norm}
{\cal {N}}_{nl}=\sqrt{2n!\over \Gamma(n+l+{3\over 2})}\ .
\end{equation}
Putting all the elements together, the wave function is then expanded
in a set of states of the form
\begin{equation}
\vert \alpha\ \rangle=
C_A\Phi \sum_{M_L}C(L\,S\,M_L\,J-M_L\,;\,J\,M)
\Psi_{_{LM_Ln_{\rho}l_{\rho}n_{\lambda}l_{\lambda}}}
\chi_{_{S\,M-M_L}}.
\end{equation}
\noindent
The entire wave function is now explicitly antisymmetric under
the exchange of quarks one and two.
\chapter{Transition Amplitude}
\label{ap:TA}
The final form of the transtition amplitude is
\begin{eqnarray} \label{ampa1}
&&M_{A \to BC}={6\gamma \over 3 \sqrt{3}}(-1)^{J_a+J_b+\ell_a+\ell_b
-1}\sum_{J_\rho,s_a,s_b}
 {\hat J}\/_\rho^2 \hat s_a {\hat S}\/_a {\hat L}\/_a \hat s_b {\hat S}\/_b 
{\hat L}\/_b\nonumber \\
&&\left \{\matrix{S_a&L_\rho&s_a\cr
               \ell_a&J_a&L_a\cr} \right \}
\left\{\matrix{L_\rho&S_\rho&J_\rho\cr
                {1 \over 2}&s_a&S_a} \right\} 
\left \{\matrix{S_b&L_\rho&s_b\cr
               \ell_b&J_b&L_b\cr} \right \}
\left\{\matrix{L_\rho&S_\rho&J_\rho\cr
                {1 \over 2}&s_b&S_b} \right\} \nonumber \\
&& (-1)^{\ell+\ell_a+J_c-L_c-S_c}{\cal F}(ABC) {\cal R}(ABC)\nonumber \\
&&\times \sum_{S_{bc}}(-1)^{s_a-S_{bc}} 
\left [\matrix{J_\rho&1/2&s_b\cr
               1/2&1/2&S_c\cr
               s_a&1&S_{bc}} \right ]
\sum_{L_{bc}} (-1)^{L_{bc}} 
               \left [\matrix{s_b&\ell_b&J_b\cr
                              S_c&L_c&J_c\cr
                              S_{bc}&L_{bc}&J_{bc}} \right ]\nonumber \\
&&\times \sum_L {\hat L}\/^2
 \left\{\matrix{s_a&\ell_a&J_a\cr
                L&S_{bc}&1} \right\}
\left\{\matrix{S_{bc}&L_{bc}&J_{bc}\cr
                 \ell&J_a&L} \right\} \varepsilon(\ell_b,L_c,L_{bc},\ell,\ell_a,L,k_0).
\end{eqnarray}
Here 
\begin{equation}
{\bf J}_a={\bf L}_a + {\bf S}_a = {\bf \ell}_a + {\bf s}_a,
\end{equation}
with
\begin{eqnarray} \label{recoup1}
{\bf L}_a&=&{\bf L}_{\lambda_a} + {\bf L}_{\rho_a}\equiv{\bf \ell}_a + {\bf L}_{\rho_a}, \nonumber \\
{\bf S}_a&=&{\bf S}_{\rho_a} + {\bf 1/2},
\end{eqnarray}
and
\begin{eqnarray}
\label{recoup2} {\bf s}_a&=&{\bf J}_{\rho_a} + {\bf 1/2}={\bf L}_{\rho_a} + {\bf S}_{\rho_a} + {\bf 1/2},
\end{eqnarray}
with similar definitions for $B$. The first four $6-j$ symbols of Eq.~(\ref{ampa1}) are necessary for transforming from the usual angular momentum basis for the baryons, given by Eq.~(\ref{recoup1}), to the basis of Eq.~(\ref{recoup2}), which is the more convenient one for evaluating the transition amplitude. $L$, $L_{bc}$ and $S_{bc}$ are internal summation variables, and ${\cal F}(ABC)$ is the flavor overlap for the decay.

The purely ``spatial'' part of the transition amplitude is
\begin{eqnarray} \label{ampa2}
&&\varepsilon (\ell_b,L_c,L_{bc},\ell,\ell_a,L,k_0)={\cal J}(A)(-1)^{L_{bc}} {1 \over 2}
{\exp{(-F^2k_0^2)} \over G^{\ell_a+\ell_b+L_c+4}} N_a N_b N_c\nonumber \\
&&\times \sum_{\ell_1,\ell_2,\ell_3,\ell_4} C^{\ell_b}_{\ell_1} C^{L_c}_{\ell_2}
C^1_{\ell_3} C^{\ell_a}_{\ell_4}\left(x-\omega_1\right)^{\ell_1}
\left(x-\omega_2\right)^{\ell_2} \left(x-1\right)^{\ell_3} 
x^{\ell_4}\nonumber \\
&&\times \sum_{\ell_{12},\ell_5, \ell_6, \ell_7, \ell_8}
(-1)^{\ell_{12}+\ell_6} 
{\hat \ell_5 \over \hat L}\left [\matrix{\ell_1&\ell_1^\prime&\ell_b\cr
                           \ell_2&\ell_2^\prime&L_c\cr
                           \ell_{12}&\ell_6&L_{bc}} \right ]
\left [\matrix{\ell_3&\ell_3^\prime&1\cr
               \ell_4&\ell_4^\prime&\ell_a\cr
               \ell_7&\ell_8&L} \right ]\nonumber \\
&&\times\left\{\matrix{\ell&\ell_{12}&\ell_5\cr
               \ell_6&L&L_{bc}} \right\} B^{\ell_{12}}_{\ell_1 \ell_2} B^{\ell_5}_{\ell \ell_{12}}
B^{\ell_6}_{\ell_1^\prime \ell_2^\prime} B^{\ell_7}_{\ell_3 \ell_4}
B^{\ell_8}_{\ell_3^\prime \ell_4^\prime}\nonumber \\
&&\sum_{\lambda,\mu,\nu} D_{\lambda \mu \nu}(\omega_1,\omega_2,x) I_\nu(\ell_5,\ell_6,\ell_7,\ell_8;L)
\left({\ell_1^\prime+\ell_2^\prime+\ell_3^\prime+\ell_4^\prime+2\mu+\nu+1 \over 2}\right)!\nonumber \\
&&\times k_0^{\ell_1+\ell_2+\ell_3+\ell_4+2\lambda+\nu}/G^{2\mu+\nu-\ell_1-\ell_2-\ell_3-\ell_4}.
\end{eqnarray}

In this expression, $N_a$ is a normalization coefficient that results from writing a single component of the wave function of $A$ as
\begin{eqnarray} 
&&\Psi_{LMn_\rho \ell_\rho n_\lambda \ell_\lambda}({\bf p}_1,{\bf p}_2,{\bf p}_3)=\eta (AA^\prime)^{3/2} \sum_m <\ell_\rho\ell_\lambda m M-m|LM> \nonumber \\
&&\times{\cal N}_{n_\rho\ell_\rho}(A^\prime p_\rho)^{\ell_\rho}e^{-{A^{\prime^2}p_\rho^2\over 2}} L_{n_\rho}^{\ell_\rho+1/2} (A^\prime p_\rho)Y_{\ell_\rho m}({\bf \Omega}_\rho)\nonumber \\
&&\times{\cal N}_{n_\lambda\ell_\lambda}(Ap_\lambda)^{\ell_\lambda}e^{-{A^2p_\lambda^2\over 2}} L_{n_\lambda}^{\ell_\lambda+1/2} (Ap_\lambda)Y_{\ell_\lambda m}({\bf \Omega}_\lambda).
\end{eqnarray}
For proper exchange symmetry among the quarks, $A^\prime={2\over\sqrt{3}}A$, and
\begin{equation} \label{prho}
{\bf p}_\rho={1\over 2}\left({\bf p}_1-{\bf p}_2\right),\,\, {\bf p}_\lambda={1\over 3}\left({\bf p}_1+{\bf p}_2-2{\bf p}_3\right).
\end{equation}
$\eta$ is a phase factor that arises from calculating the Fourier transform of the configuration space wave functions, and has the value
\begin{equation}
\eta=(-i)^{2n_\rho+2n_\lambda+\ell_\rho+\ell_\lambda}.
\end{equation}
With these definitions, $N_a=A^{\ell_\lambda+3/2}{\cal
N}_{n_\lambda\ell_\lambda}$, with ${\cal N}_{n\ell}$ previously defined
in~\ref{Norm}, $L_n^{\ell+1/2}$ in~\ref{Laguerre}, while the $Y_{\ell m}$ are the usual spherical harmonics.

${\cal J}$ is a Jacobian factor needed to convert from the basis used in evaluating the space factor $\varepsilon$ in Ref.~\cite{RoSB:1992}, to the basis used in the evaluation of the wave functions used for explicit calculation of the decay amplitudes. The wave functions of Ref.~\cite{CaIs:1986} use
\begin{equation} 
{\bf p}_\rho^\prime={1\over\sqrt{2}}\left({\bf p}_1-{\bf p}_2\right),\,\, {\bf p}_\lambda^\prime={1\over\sqrt{6}}\left({\bf p}_1+{\bf p}_2-2{\bf p}_3\right),
\end{equation}
so that both the Jacobian factor mentioned above, as well as a redefinition of the gaussian parameters of the wave functions, are required in order to use the wave functions of Ref.~\cite{CaIs:1986} with the above expression for the decay amplitude.

The factor ${\cal R}$ of Eq.~(\ref{ampa1}) is obtained as the overlap
of the wave functions in the $\rho$ coordinates in the initial and
final baryon. Since in the model used here quarks $1$ and $2$ are spectators ($\ell_{\rho_a}=\ell_{\rho_b}$, $S_{\rho_a}=S_{\rho_b}$, $J_{\rho_a}=J_{\rho_b}$), and the basis is fully orthogonalized ($\alpha$ is the same in the initial and final baryons, so that $n_{\rho_a}=n_{\rho_b}$), this overlap is always unity. In addition, this means that the Jacobian discussed above is only necessary for the transformation in ${\bf p}_\lambda$.

The $\sum_{\lambda,\mu,\nu} D_{\lambda \mu \nu}(\omega_1,\omega_2,x) I_\nu(\ell_5,\ell_6,\ell_7,\ell_8;L)$ term arises from writing (here ${\bf q}_a\equiv{\bf p}_{\lambda_a}$, with a similar definition for the daughter baryon)
\begin{eqnarray}
&&L_{n_{\lambda_a}}^{\ell_a}e^{-A^2q_a^2/2} L_{n_{\lambda_b}}^{\ell_b}e^{-B^2q_b^2/2} L_{n_c}^{L_c}e^{-C^2q_c^2/2}\nonumber \\
&&\equiv\sum_{\lambda,\mu,\nu} D_{\lambda \mu \nu}(\omega_1,\omega_2,x) e^{-A^2q_a^2/2} e^{-B^2q_b^2/2} e^{-C^2q_c^2/2}.
\end{eqnarray}
When the substitutions ${\bf q}_a=x{\bf k} +{\bf q}$, ${\bf
q}_b=(x-\omega_1){\bf k} +{\bf q}$, ${\bf q}_c=(x-\omega_2){\bf k}
+{\bf q}$ are made, and the integrals over ${\bf k}$ and ${\bf q}$ are
evaluated, the expression above results. The full form of the
$D_{\lambda\mu\nu}$ does not provide additional information so it omitted.

In Eqs.~(\ref{ampa1}) and (\ref{ampa2}),
\begin{equation}
\left [\matrix{a&b&c\cr
                 d&e&f\cr
                 g&h&i} \right ]
= \hat c \hat f \hat g \hat h \hat i \left\{\matrix{a&b&c\cr
                 d&e&f\cr
                 g&h&i} \right\}
\end{equation}
where $
\left\{\matrix{a&b&c\cr
                 d&e&f\cr
                 g&h&i} \right\}$
is the $9-j$ symbol, and ${\hat J}\/= \sqrt{2J+1}$.

In Eq.~(\ref{ampa2})
\begin{eqnarray}
x&=&\left(B^2\omega_1+C^2\omega_2+f^2 \right)
\left(A^2 +B^2 +C^2 +f^2 \right)^{-1},\nonumber \\
F^2&=&{1 \over 2}\left[A^2x^2 +B^2\left(x-\omega_1\right)^2+C^2\left(x-\omega_2\right)^2+f^2(x-1)^2\right],\nonumber \\
G^2&=&{1 \over 2}(A^2+B^2+C^2+f^2).
\end{eqnarray}
$\omega_1$ and $\omega_2$ are ratios of various linear combinations of quark masses. In general,
\begin{equation}
\omega_1={m_1+m_2\over m_1+m_2+m_4},\,\,\omega_2={m_3\over m_3+m_4},
\end{equation}
where the subscripts refer to the quark labels shown in
Figure~\ref{figB1}. In addition,
\begin{eqnarray}
C^\ell_{\ell_1}&=&\sqrt{{4\pi (2\ell+1)! \over (2\ell_1+1)!
[2(\ell-\ell_1)+1]!}},\nonumber \\
B^\ell_{\ell_1 \ell_2} &=&{(-1)^\ell \over \sqrt{4\pi}} {\hat \ell}\/_1
{\hat \ell}\/_2 \left(\matrix{\ell_1&\ell_2&\ell\cr
                 0&0&0} \right),
\end{eqnarray}
and $\ell_1^\prime=L_b-\ell_1$, $\ell_2^\prime=L_c-\ell_2$,
$\ell_3^\prime=1-\ell_3$, $\ell_4^\prime=L_a-\ell_4$ and the geometric factor $I_\nu$ is
\begin{eqnarray}
&&I_{2p}(\ell_5,\ell_6,\ell_7,\ell_8;L)=(-1)^L(2p)! {\hat \ell}_5 {\hat \ell}_6 {\hat \ell}_7
{\hat \ell}_8\nonumber \\
&&\times \sum_{\lambda=0}^p {4^\lambda (4\lambda+1)(p+\lambda)! \over
(2p+2\lambda+1)!(p-\lambda)!} 
\left(\matrix{2\lambda&\ell_5&\ell_7\cr
              0&0&0}\right)
\left(\matrix{2\lambda&\ell_6&\ell_8\cr
              0&0&0}\right)
\left\{\matrix{\ell_5&\ell_6&L\cr
              \ell_8&\ell_7&2\lambda}\right\},\nonumber \\
&&I_{2p+1}(\ell_5,\ell_6,\ell_7,\ell_8;L)=2(-1)^{L+1}(2p+1)! {\hat \ell}_5 {\hat \ell}_6
{\hat \ell}_7 {\hat \ell}_8\nonumber \\
&&\times \sum_{\lambda=0}^p {4^\lambda (4\lambda+3)(p+\lambda+1)!
\over
(2p+2\lambda+3)!(p-\lambda)!} 
\left(\matrix{2\lambda+1&\ell_5&\ell_7\cr
              0&0&0}\right)\nonumber \\
&&\times\left(\matrix{2\lambda+1&\ell_6&\ell_8\cr
              0&0&0}\right)
\left\{\matrix{\ell_5&\ell_6&L\cr
              \ell_8&\ell_7&2\lambda+1}\right\}.
\end{eqnarray}
\chapter{Computational Methods}
\label{ap:COMP}
There are two main computational elements required to accomplish the 
calculation in this work. Recall the main equation of Chapter~\ref{loops}
\begin{equation}
\label{loop_eq_c}
Re[\Sigma_B(E)] = \sum_{B'M} {\cal P}\int_0^\infty \frac{k^2dk
{\cal M}_{BB'M}^{\dagger}(k){\cal M}_{BB'M}(k)}{E-\sqrt{M_{B'}^2 + k^2} -
\sqrt{m_M^2 + k^2}},
\end{equation}
where ${\cal M}_{BB'M}(k)$ is an expression for the momentum dependent
vertex of the strong decay $B \rightarrow B'M$, and ${\cal P}$
indicates that only the real part of the integral is evaluated by
principal-part integration. How these components are computed is
described in general terms in what follows, along with a sequence of
how they are combined to produce the results presented in
Chapter~\ref{ResI}.
\section{Strong Decay Vertices}
The momentum dependence of the strong decay vertex can be obtained
from the analytical form of the strong decay matrix element $\langle
B'M|T|B \rangle$, with $T$ representing the $\tp0$ decay operator. The
method used to evaluate these matrix elements is based on the work of
Ref.~\cite{RoSB:1992} and Ref.~\cite{CaRo:1993}, and several of the
angular momentum techniques can be found in Ref.~\cite{VMK}.

The symbolic manipulator {\it Maple} was used to compute all the
components involved in calculating each decay matrix element for a
large set of decays $B\rightarrow B'M$. Subroutines were built to
independently calculate each component so that it could be
individually tested before being integrated into a higher level
subroutine, and so that it could be totally portable. Basic procedures
such as those to analytically calculate Clebsch-Gordan, $6-j$, or
$9-j$ coefficients were implemented, along with generators of
spherical harmonics and associated Laguerre polynomials. Then another
series of algorithms were devised and coded to handle each component
of Eq.~\ref{ampl} as defined in Appendix~\ref{ap:TA}. The very large
number of nested summations involved required careful and thorough
testing as the trouble-shooting became more complex with each `layer'.

Due to the symmetry of the wave functions, in this $\tp0$ model decays
involving the first or second quark are considered separately from
those involving the third quark of the decaying baryon. However, the
two cases are related by a set of coefficients that, in effect,
project a set of basis states onto another, or equivalently rotate a
set of coordinates. The matrix elements can then be calculated in two
different bases, and then combined after one set of results is
transformed. These coefficients were adapted for this work from
Moshinsky~\cite{Mo:1959,Mo:1967,Mo:1969} brackets, coded, and then
subsequently mass-produced with the results stored in analytical
form. Once that was accomplished, it became possible to rapidly
analytically compute matrix elements for decays from {\bf any} initial
baryon to {\bf any} baryon-meson state, making this code a very
versatile and powerful tool.

Since {\it Maple} is a powerful but slow program, and large parts of
the analytical decay matrix elements are common for specific sets of
quantum numbers, matrices of decay matrix elements were created
between sets of hadron quantum numbers based on the expansion of the
baryon wave functions. For example, matrices were built for decays
from $J^P=\thalf^+$ baryons ({\it e.g.}~the $\Delta$ baryon) to
$J^P=\half^+$ baryons ({\it e.g.}~the nucleon) and a pseudo-scalar
meson ({\it e.g.}~the pion). Since the baryon wave functions used in
this project are expanded to the $N=6$ level (for positive-parity
states and $N=7$ for negative-parity states) and thus have an average
of about 100 components, for this example one matrix is 78 x 50, the
other 152 x 100, with each element corresponding to a strong decay
between a pair of basis substates. The process is repeated for decays
involving vector mesons ({\it e.g.}~the rho). Given the number of
intermediate baryon-meson states included in this work, it is an
understatement to say that this was a computer-intensive endeavor. It
took several months to complete, using many nodes on the FSU Computer
Cluster. Matrices were finally tested by using them to reproduce a
large variety of published decay
amplitudes~\cite{CaRo:1993,CaRo:1994,CaRo:1998}.

The reward for this extensive calculation is the versatility of the
code. Not only can it be used in the type of project described in this
dissertation but, as mentioned above, it can also be applied to obtain
any strong decay amplitude within the limits of the $\tp0$
model. Since the decay matrix elements are stored in analytical form,
changes in the value of any parameter can easily be handled without
having to recalculate anything, and since wave function expansion
coefficients are independent of the decay matrix elements, they can
also be changed without affecting this part of the code.
\section{Numerical Integration}
Since the integrand of Eq.~\ref{loop_eq_c} can become extremely
complex, it is not practical, and often not possible, to try to do the
integration analytically, so numerical integration schemes were
therefore used. Since both real (the initial baryon's energy is above
the threshold for production of the intermediate baryon-meson pair)
and virtual (below threshold) decays are encountered, two different
routines were used. In the case of virtual decays the integrand is
always real, so a Gauss-Laguerre quadrature routine was used to perform
the integration. Above threshold decays imply the presence of a pole,
therefore a combination of Gauss-Laguerre and Gauss-Legendre type
quadratures were used to integrate symmetrically around the pole and
evaluate the principal part of Eq.~\ref{loop_eq_c}. 

The integration routines were developed using the {\it C} programming
language, and used modified versions of pre-coded numerical algorithms
from Ref.~\cite{NRC}.
\section{Overall Scheme}
The numerator of Eq.~\ref{loop_eq_c} is obtained via a {\it Maple}
routine which performs the matrix algebra required to combine the
matrices of decay matrix elements with the appropriate vectors of wave
function expansion coefficients. The result is an analytical
expression as a function $k$ which is then translated into {\it C}
code by {\it Maple}. The result is then used by the integration
routine. 

During the matrix algebra part of the calculation, it was found that
the amount of time required to process analytical matrices ranging in
size from 150 MB (for decays involving pseudo-scalar mesons) to well
over 650 MB (for decays involving vector mesons) could limit the
number of intermediate baryon-meson states included. Since a crucial
ingredient of the calculation is the large extent of the sum over
intermediate states, an additional routine was implemented to
`preprocess' the matrices by assigning numerical values to the
parameters that were common to a set of initial states (based on a
given set of baryon wave functions), which were then stored in new
matrices for use when needed. This extra step, added at the cost of a
few days of processing for each new set of matrices, cut the final
matrix algebra time by a factor of roughly 60\%, making it possible to
compare multiple sets of results within a reasonable time frame.

For each initial baryon and each baryon-meson combination in the sum
over intermediate states ($\sum_{B'M}$), the associated integral is
evaluated for 200 different values of $E$ over the energy range being
considered. These values are then tabulated and then used to produce
graphs such as those presented in Chapter~\ref{ResI}.

Throughout this process, liberal use is made of scripts (both in {\it
Maple} and {\it Unix}) to handle the large number of intermediate states
and automate the process as much as possible.


\bibliography{thesis}

\end{document}